\documentclass[aps,pra,showpacs,twocolumn,superscriptaddress]{revtex4-1}

\usepackage[squaren,Gray]{SIunits}
\usepackage{multirow}
\usepackage{graphicx}

\usepackage{amsmath,amssymb,bbm}
\usepackage{dcolumn}
\usepackage{bm}
\usepackage{array}
\usepackage{color}
\usepackage{braket}
\usepackage{color}
\usepackage{booktabs}

\newcommand{\rB}{\mbox{\boldmath$r$}}
\newcommand{\RB}{\mbox{\boldmath$R$}}

\AtBeginDocument{
\heavyrulewidth=.08em
\lightrulewidth=.05em
\cmidrulewidth=.03em
\belowrulesep=.65ex
\belowbottomsep=0pt
\aboverulesep=.4ex
\abovetopsep=0pt
\cmidrulesep=\doublerulesep
\cmidrulekern=.5em
\defaultaddspace=.5em
}

\begin{document}


\title{Universality of fold-encoded localized vibrations in enzymes}



\author{Yann Chalopin}
\affiliation{Laboratoire d’Energétique Macroscopique et Moléculaire, Combustion (EM2C), CentraleSupélec, CNRS, 91190 Gif-sur-Yvette, France}

\author{Francesco Piazza} 
\affiliation{Centre de Biophysique Moléculaire (CBM) CNRS UPR4301 $\&$ Université d’Orléans, Orléans 45071, France}

\author{Svitlana Mayboroda}
\affiliation{School of Mathematics, University of Minnesota, Minneapolis, Minnesota 55455, USA}

\author{Claude Weisbuch}
\affiliation{Laboratoire de Physique de la Matière Condensée, Ecole Polytechnique, CNRS, 91128 Palaiseau, France}
\affiliation{Materials Department, University of California, Santa Barbara, California 93106, USA}

\author{Marcel Filoche}
\affiliation{Laboratoire de Physique de la Matière Condensée, Ecole Polytechnique, CNRS, 91128 Palaiseau, France}



\date{\today}

%
%
%
%
%
\begin{abstract}
Enzymes speed up biochemical reactions at the core of life by as much as 15~orders of magnitude. Yet, despite considerable advances, the fine dynamical determinants at the microscopic level of their catalytic proficiency are still elusive. In this work, we use a powerful mathematical approach to show that rate-promoting vibrations in the picosecond range, 
specifically encoded in the 3D protein structure, are {\em localized} vibrations optimally coupled to the chemical reaction coordinates at the active site. The universality of these features is demonstrated on a pool of more than 900 
enzyme structures, comprising a total of more than 10,000 experimentally annotated catalytic sites. 
Our theory provides a natural microscopic rationale for the known subtle structural compactness  
of active sites in enzymes.  
%
%
\end{abstract}


\maketitle


%
\section{Introduction}

\noindent  The intricate networks of metabolic cascades that power living organisms ultimately rest on the exquisite ability of enzymes to increase the rate of chemical reactions by many orders of magnitude. However, despite a large body of evidence accumulated over the past two decades in favor of the highly dynamical nature of proteins, the question whether protein motions such as conformational changes and finer (and faster) reorganization dynamics play a role in enzyme catalysis remains widely debated~\cite{Nagel2009}.\\
\indent Although many molecular machines contain intrinsically disordered domains~\cite{Oldfield2014}, the 3D fold is central to enzyme functioning. In particular, increasing evidence is accumulating in the literature in favor of the existence of specific fold-encoded motions believed to be optimally coupled to the chemical reaction coordinate(s)~\cite{Zinovjev2017, Kale2008, Agarwal2005, Antoniou2001, Hay2012, Luk2013, Nagel2009}. These motions typically correspond to localized vibrations of the protein scaffold that  contribute to the catalytic reaction, i.e. modes that, if impeded, would lead to a deterioration of the catalytic efficiency~\cite{Nagel2009}. The existence and importance of such localized, shape-specific motions, coined rate-promoting vibrations (RPV)~\cite{Antoniou2001} is backed by many computational  and experimental studies~\cite{Kale2008, Pudney2009, Heyes2009, Heyes2011, Henzler-Wildman2007c, Saen-Oon2008, Masterson2010, Agarwal2002}, beginning with the pioneering ideas by McClare on the functional role of non-equilibrium localized motions in muscle contraction~\cite{McClare1972}. The role of RPVs in enzymes has been highlighted for the tunneling reaction coordinate in lactate dehydrogenase (LDH)~\cite{Chen2018, Dzierlenga2016, Quaytman2007}. Promoting modes in Purine Nucleosidase phosphorylase (PNP) have also been explored more recently~\cite{Harijan2017}. Interestingly, evidence for the existence of promoting vibrations coupling directly to the reaction coordinate in enzyme-catalyzed hydrogen transfer reactions has also been gathered from the temperature dependence of kinetic isotope effect (KIE)~\cite{Arcus2015}. More generally, the key rate-promoting role of fluctuations in the region of the active site has been established on rigorous quantum mechanical grounds in the 1990s by Bruno and Bialek for enzymatic hydrogen transfer~\cite{Bruno1992}. Yet, despite the broad set of evidence for specific dynamical effects in enzymes-catalyzed reactions, a universal demonstration of the existence of RPVs in enzymes that could explain how specific vibrations at the active site contribute to increase the reaction rate is still lacking.\\
\indent To tackle the problem of assessing the role of vibrations in the catalytic efficiency of enzymes, it is essential to understand that in general protein motions play a rather diverse and subtle role over a wide range of timescale and distances~\cite{McCammon:1987aa}. The longest times, which correspond to conformational changes of the protein, are in the ms-s range~\cite{Wolf-Watz2004} and are generally believed not to be directly coupled to the enzymatic catalytic step, as most enzymes have turnover rates in the $10^{3}$ s$^{-1}$ ballpark~\cite{Nagel2009}. The matter is subtler for allosteric transitions (i.e., action at a {\em distance})~\cite{Changeux2005}, and slow conformational sampling, occurring in the ms-s timescale too, with many studies advocating a variable degree of coupling of those motions to the chemical step~\cite{Hammes2002, Gerhart1968}, including the key advances brought about by single-molecule enzymology~\cite{English2006, Xun:1998aa}. Faster conformational sampling in the ns-ms and faster reorganization motions of the active sites in the ps-ns range are commonly accepted to play an important role in shaping the kinetic behavior of many enzymes, such as alcohol dehydrogenase~\cite{Liang2004} and methylamine dehydrogenase~\cite{Basran1999}, as most clearly revealed by the pioneering studies on the role of protein motions in hydrogen tunneling in soybean lypoxygenase-1~\cite{Liang2004, Knapp2002}.\\
\indent Quantum-mechanical tunneling in hydrogen transfer at room temperature was first demonstrated in 1989 in a seminal paper on alcohol dehydrogenase~\cite{Cha1989}. In particular, this discovery revealed the tremendous power of  kinetic isotope effects studies to investigate the direct coupling of fast vibrational modes localized at the active site to the catalytic step~\cite{Klinman2013}. The general surprising finding is that the KIE is largely temperature-independent in many native enzyme systems~\cite{Knapp2002, Pudney2009, Basran1999}. This is usually interpreted as the blueprint of an optimal structural 
{\em compactness} at the active site, where reaction partners are kept tight in the optimal geometry that underlies the catalytically competent atomic arrangement. This fact perfectly rhymes with the known reports that active sites tend to lie in the stiffest regions of enzyme structures~\cite{Sacquin-Mora2007, Juanico2007, Aubailly2015} and that a subtle balance of rigidity and some specific flexibility are implied in enzyme catalysis~\cite{Kamal:2012aa, Guo:2012aa}.\\
\indent Taken together, the above facts lead to an emerging picture where enzymes feature highly compact, pre-organized active sites. These represent structurally competent catalytic precursors that are generically modulated through slow conformational sampling at the level of the whole structure, but more finely and specifically regulated by specific rate-promoting vibrations that couple directly to the reaction coordinate(s). Hydrogen tunneling kinetics provides the perfect grounds for illustrating these ideas. There is now a wide consensus that donor-acceptor distances (DAD) at the active site for enzymes that catalyze the transfer of some hydrogen species are modulated with sampling frequencies in the \unit{50-300}{cm^{-1}} range~\cite{Klinman2013}, corresponding to motions in the ps-ns range. These RPVs provide optimal compression along the DAD, thus enhancing the tunneling rate through a vibrationally assisted mechanism~\cite{Klinman2013, Bruno1992}. In other words, fast conformational sampling along the DAD is optimal in the substrate-bound conformation, which generates active-site compression leading to favorable close approach between donor and acceptor atoms~\cite{Klinman2013} on timescales slower than tunneling times (fs).\\
\indent In this paper, we take one step forward and show that fold-specific, localized vibrations enforcing dynamical compression at the active site are a universal feature of enzymes. This suggests that enzymes structures have evolved as optimally designed {\em mechanical transducers} of vibrational energy mediated by 
RPV patterns~\cite{Heyes2011}. The article is organized as follows: first, we introduce the localization landscape (LL), a novel and powerful mathematical tool which we use here to predict the spatial distribution of energy in proteins modeled by the Elastic Network Model (ENM)~\cite{Atilgan2001}. The implementation of the LL is illustrated for a specific example, the well-documented case of LDH, before reporting the results of a systematic study of the correlation between active sites and localized vibrations on a sample of about 1,000 enzymes (corresponding to more than 10,000 annotated actives sites).


\section{Materials and Methods}

\begin{figure}[h!]
\includegraphics[width=\columnwidth]{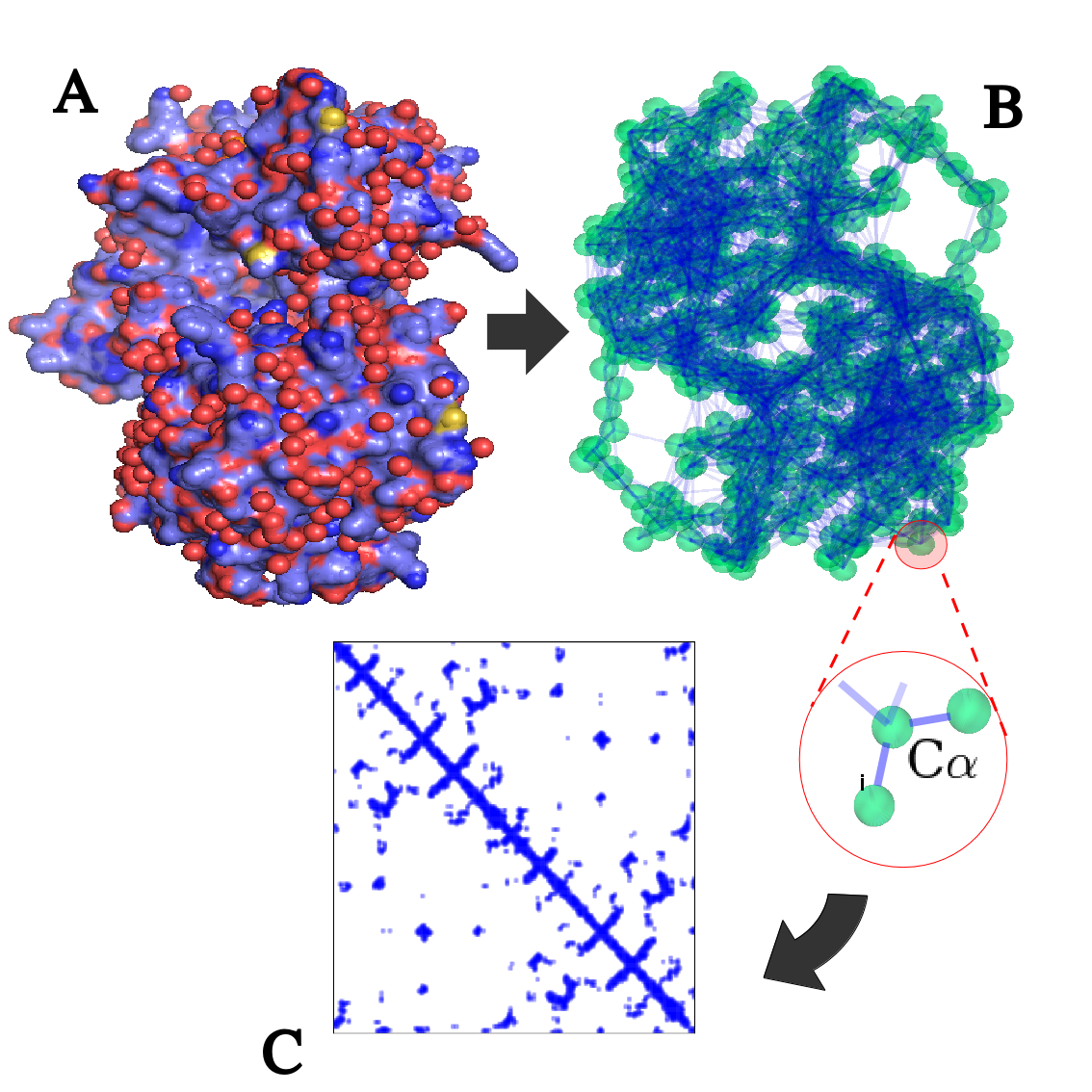}
\caption{\textbf{Coarse grained model.} \textbf{A}: All-atom view of the L-Lactate Dehydrogenase dimer (LDH, PDB id: 1I0Z). \textbf{B}: Elastic Network Model deduced from the tertiary structure. \textbf{C}: Sparsity pattern of the $3N \times 3N$ force constant matrix $\mathbb{H}_{ij}^{\alpha\beta}$ used to compute the localization landscape (see Appendix~\ref{App:ENM}). Each non-vanishing term is represented by a blue spot. In our case of uniform spring constant and sharp cutoff coupling, this matrix is a direct representation of the connectivity pattern among residues.}
\label{fig:coarse-grained}
\end{figure}

\noindent  In order to investigate the topological origin of vibrational modes related to the active-site reorganization  
in the specific timescale of interest (100 $cm^{-1}$), we adopt a coarse-grained elastic-network model (ENM)~\cite{Tirion1996, Atilgan2001, Juanico2007} (see Appendix~\ref{App:ENM}). This model reduces each protein to a collection of beads and springs that interact according to a unique, fold-encoded connectivity pattern. In our case, the beads correspond to the amino acids centered at the $C_\alpha$ carbon of the tertiary structure (Fig.~\ref{fig:coarse-grained}A). Enzymes are therefore seen as a set of coupled harmonic oscillators (Fig.~\ref{fig:coarse-grained}B). As it is well known~\cite{Bahar2005}, the local connectivity of each each amino acid is reflected in the sparsity pattern of the force constant matrix (Fig.~\ref{fig:coarse-grained}C). This connectivity controls the localization pattern of high-frequency modes ($\mathcal{O}(10^2)$ cm$^{-1}$ in $C_\alpha$-based ENM schemes).\\
%
%
\indent The main idea of this paper is to use a novel mathematical tool, coined localization landscape (LL), to decipher the subtle structure-dynamics-function relation in enzymes. The LL, which rests on a universal theory of wave localization, unveils the localization pattern of standing waves in complex or disordered media~\cite{Filoche2012}, and is extended here to the case of protein vibrations (see Appendix~\ref{App:ENM}). Bypassing the need to compute the full set of normal modes, the LL is a real-valued function computed at each site of the ENM network by solving a simple linear system based on the force constant matrix (see Appendix~\ref{App:LL}). This LL provides the essential information about the interplay between the complex protein shape and the propagation of microscopic vibrations. In particular, the ``valleys'' of the LL delineate the main regions of existence of the large-amplitude localized vibrations, thus yielding an effective functional partition of the molecule structure. In addition, the local maxima of the LL identify the most localized vibrating areas or ``hot spots'', while the corresponding values of the LL at these hot spots are very good predictors of the associated vibration frequencies~\cite{Lefebvre2016, Arnold2018} (see also appendix B and more specifically Fig.~\ref{fig:annex}). We emphasize here that the LL is about 50 times faster to compute than solving the full eigenvalue problem (see Table~\ref{tab:compare} in Appendix~\ref{App:compeff}).\\
\indent The LL reveals that the molecular architecture of enzymes seems so designed as to concentrate high-frequency vibrations  within a few domains, as it has been pointed out in previous studies~\cite{Aubailly2015, Yang2005, Lyra2015, Sacquin-Mora2007}. Moreover, the LL affords considerable new insight into how the localization pattern also segments the molecular scaffold into nearly vibrationally independent (i.e., uncoupled) clusters of amino acids. Although this work focuses on enzymes, the localization property seems to remain general for every protein. 

\section{Results}

\begin{figure}
\centering
\includegraphics[width=\columnwidth]{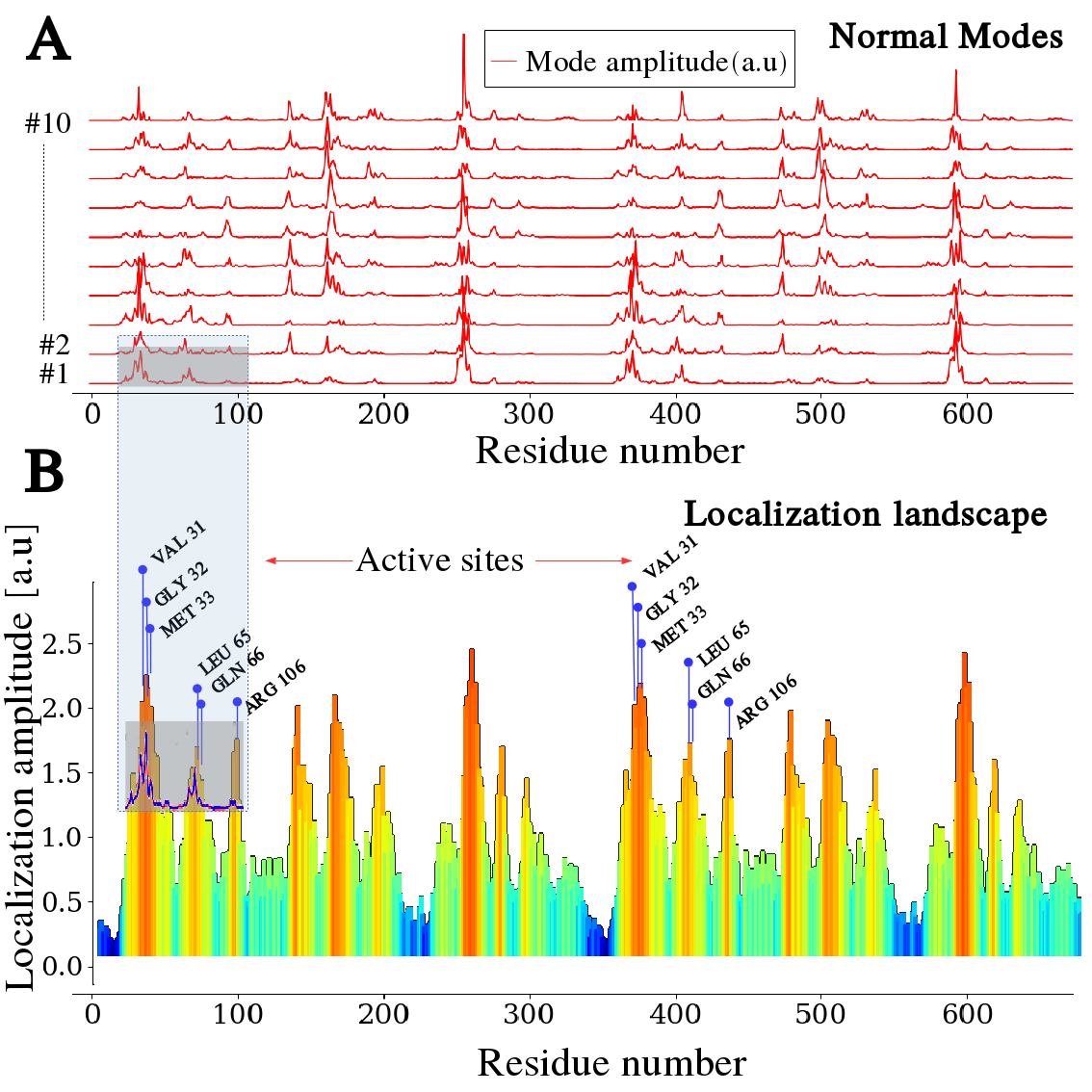}
\caption{\textbf{Comparison between Normal Modes and LL computation for human L-Lactate Dehydrogenase dimer (LDH, PDB id: 1I0Z).} \textbf{A}:~Wave localization is visualized by plotting the 10~highest-frequency normal modes. Their frequencies range from 94.2 to \unit{99.3}{cm^{-1}}. \textbf{B}:~The LL (u) is drawn along the protein backbone. Catalytic sites, shown explicitly alongside the LL, clearly lie very close to the LL maxima, corresponding to the sites of highest localization (hot spots).}
\label{fig:LDH-modes-LL}
\end{figure}

\begin{figure}
\centering
\includegraphics[width=\columnwidth]{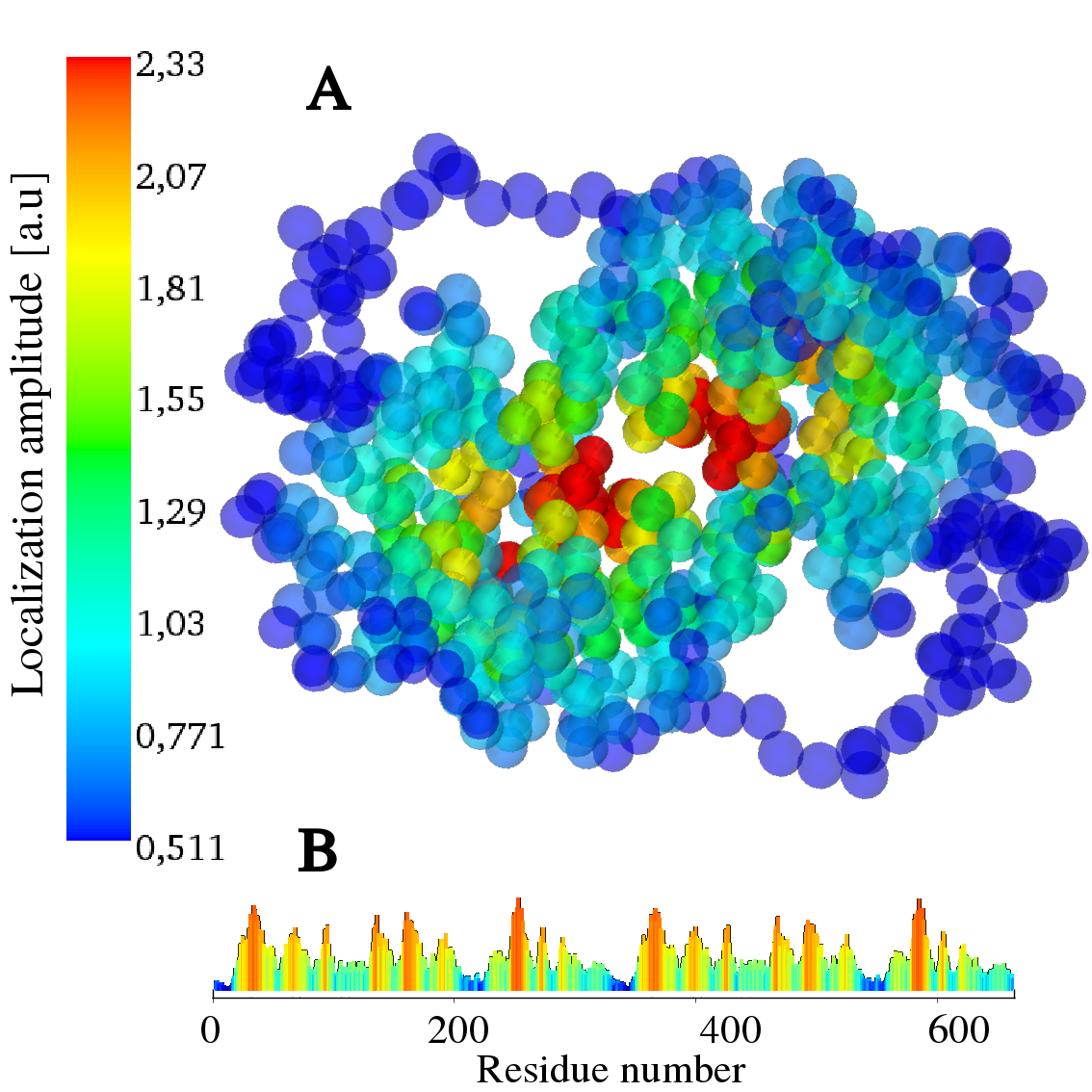}
\caption{\textbf{3D LL for human L-Lactate Dehydrogenase dimer.} The 3D LL is shown by color-coding the 3D~coarse-grained scaffold according to the amplitudes of the LL depicted in Fig.~\ref{fig:LDH-modes-LL}B. We observe here that peaks (hot spots) of the localization landscape that appear distant when plotted along the backbone chain are in fact found around the same spatial location (the red spots). Wave localization is thus predicted to occur within two distinct 3D domains lying at the center of the molecule: these domains host the catalytic activity.}
\label{fig:LDH-3DLL}
\end{figure}

\begin{figure}
\centering
\includegraphics[width=\columnwidth]{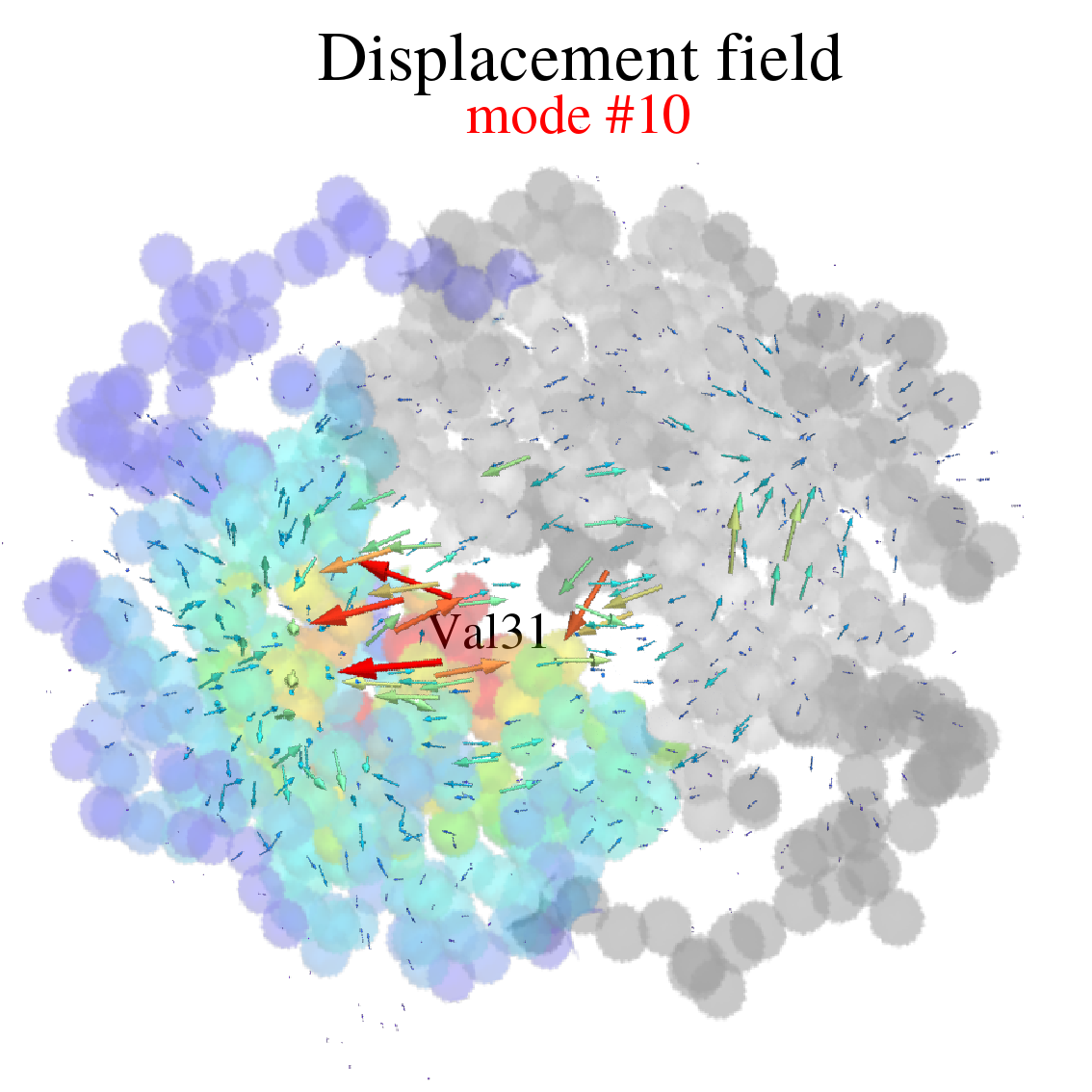}
\caption{\textbf{A rate-promoting vibration (RPV) in the L-Lactate Dehydrogenase dimer (LDH).} Localization landscape color-coded on the coarse-grained structure with a close-up of the compression field corresponding to the vibrational eigenvector \#10 (frequency \unit{94.17}{cm^{-1}}) along the reaction coordinate: residues VAL-31, GLY-32, MET-33 compress towards ARG~106. The localized eigenmode \#10 corresponds to the rate-promoting vibrations found in Ref.~\cite{Quaytman2007}.}
\label{fig:LDH-3D-compression}
\end{figure}

\begin{figure}
\centering
\includegraphics[width=\columnwidth]{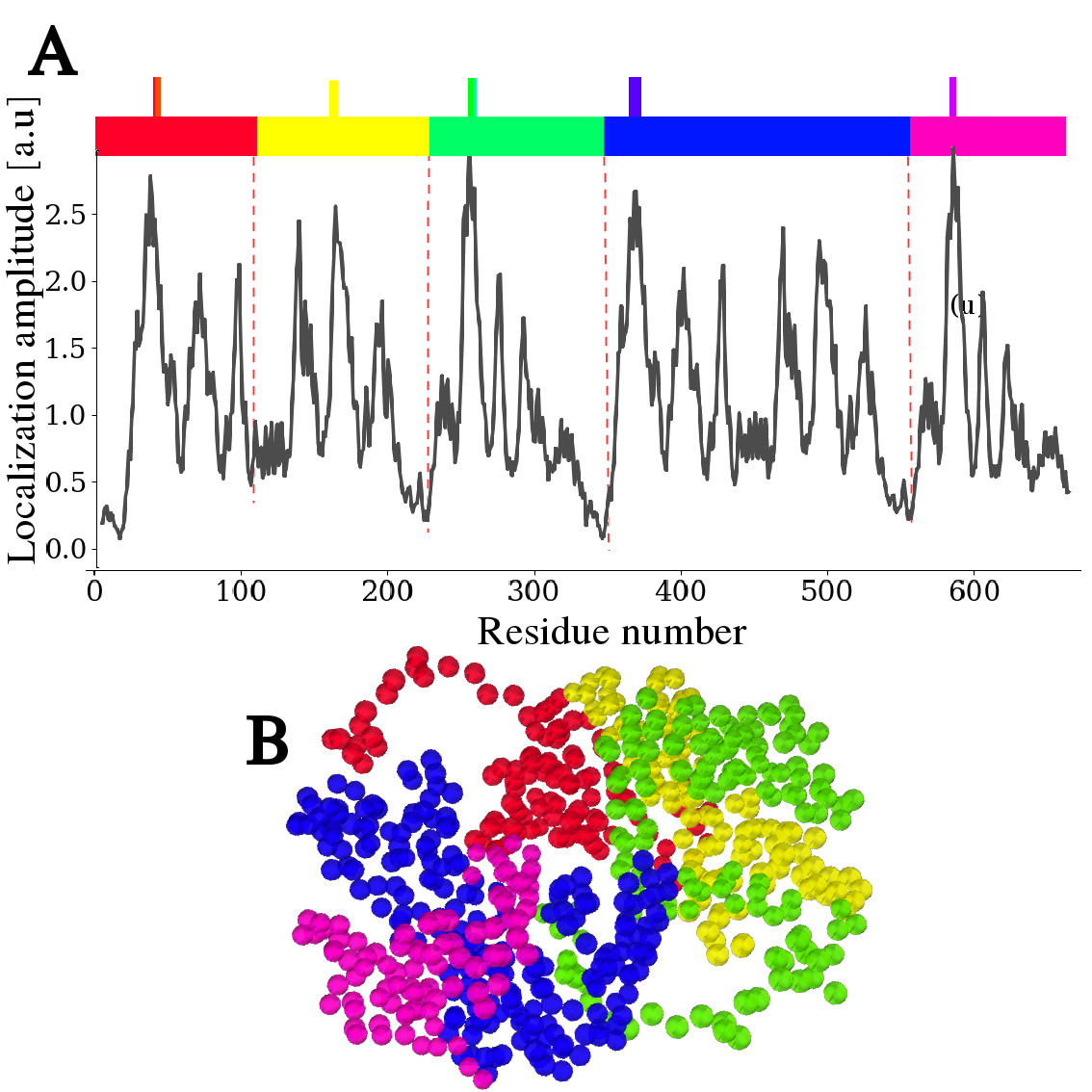}
\caption{\textbf{Localization and functional domains.} \textbf{A}: Partitioning of the molecule obtained from the LL. On the landscape plotted on the backbone chain, one selects the 4 highest local maxima (marked by a spike on the color bar) separated by the 4 lowest local minima (marked by the dotted lines). In the LL theory, each domain can be seen as a local harmonic oscillator, weakly coupled to the others. \textbf{B}: The partitioning of the LDH obtained in frame B, plotted on the tertiary structure, exhibits distinct spatial domains. }
\label{fig:LDH-domains}
\end{figure}

\begin{figure}[h!]
\centering
\includegraphics[width=\columnwidth]{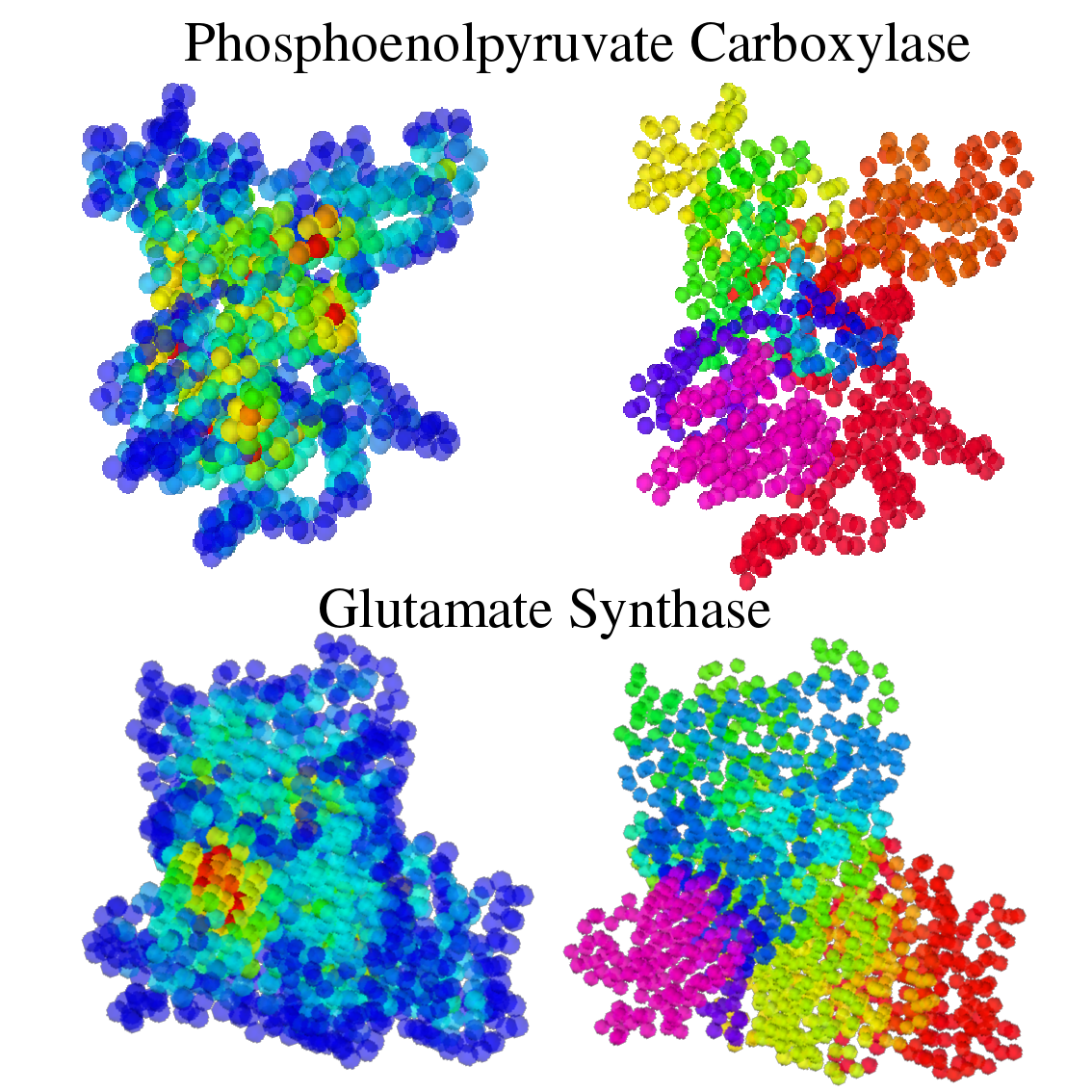}
\includegraphics[width=\columnwidth]{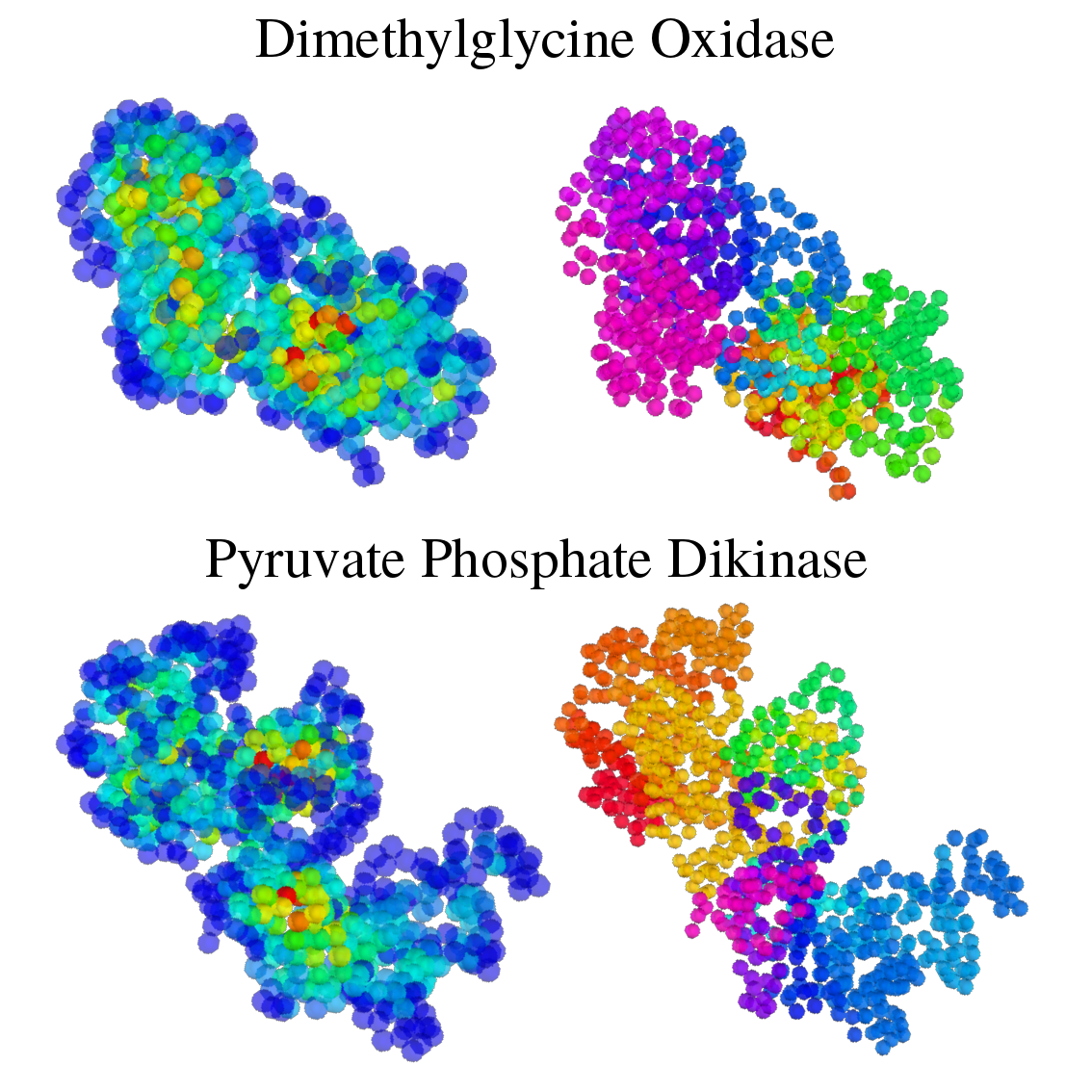}
\caption{\textbf{Domains in other enzymes.} The partitioning procedure is illustrated for four enzymes. 
The clustering of the enzymes into vibrationally independent subregions is a general feature.}
\label{fig:partition-4enzymes}
\end{figure}

\begin{figure}
\centering
\includegraphics[width=\columnwidth]{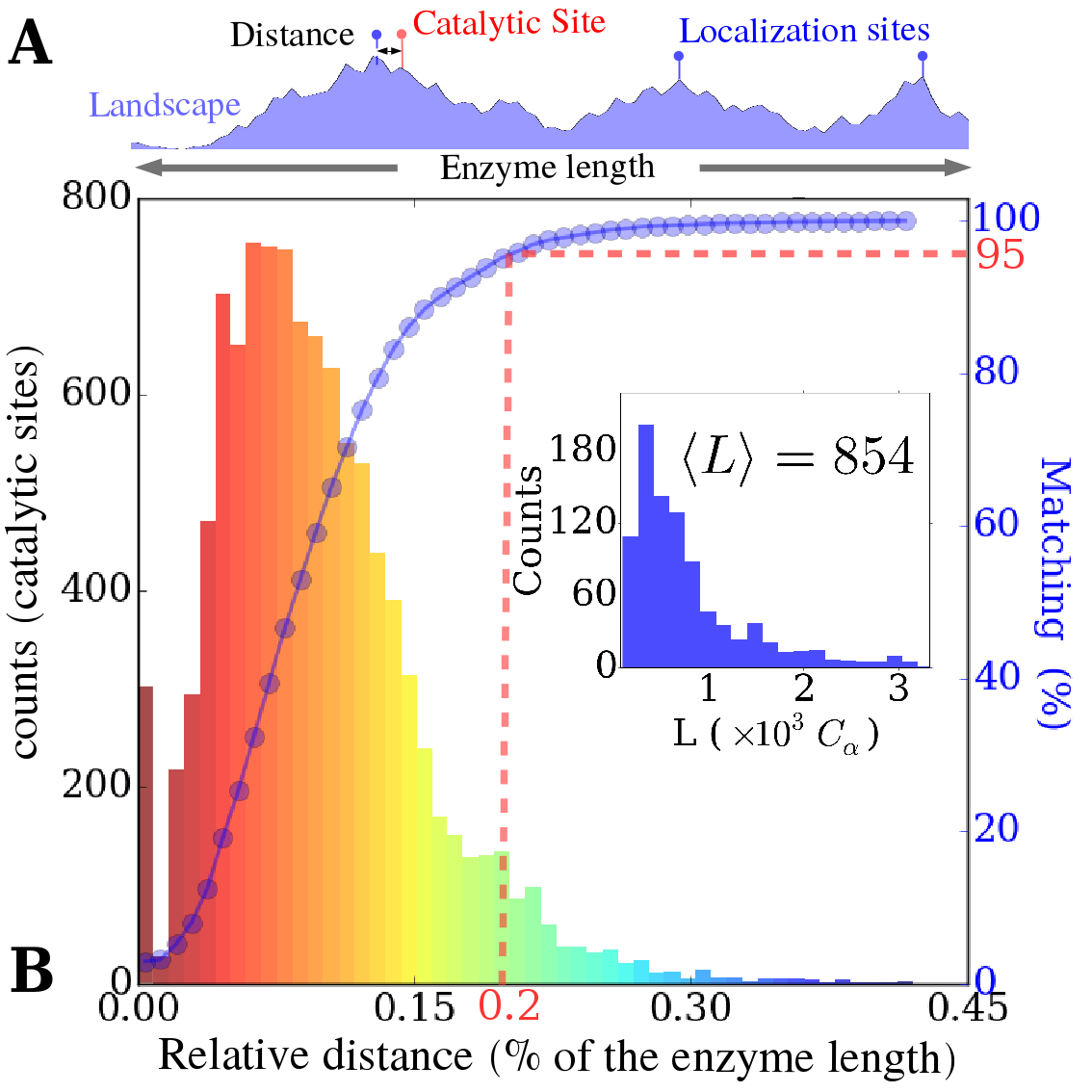}
\caption{\textbf{Proximity score for 10,566 annotated catalytic sites (933 enzymes) from the catalytic site atlas}~\cite{Porter2004}, gauging the match between a functional site and a main localization hot spot. \textbf{Frame~A}: The relative distance is scored by taking the shortest distance between catalytic sites and the main localization hot spots, divided by the chain length. \textbf{Frame~B}: Main histogram. 95\% of active sites are found at one of the highest localization spots with an error smaller that 0.2\% of the enzyme length along the chain. Inset: Size distribution.}
\label{fig:proximity}
\end{figure}

\noindent The case of LDH is presented here as a paradigmatic example to illustrate the insight offered by our method. As a comparison, we first compute all the normal modes (NM) by brute-force diagonalization of the dynamical matrix. The patterns of the highest-frequency NMs (Fig.~\ref{fig:LDH-modes-LL}A) reveal that they are highly confined to some very specific residues. We then compute the high-frequency LL of the enzyme (indicated by $u$ in Fig.~\ref{fig:LDH-modes-LL}B). The most interesting property of the LL appears when comparing it to the catalytic structure of the enzyme, characterized by the locations of the known active sites of LDH (VAL-31, GLY-32, MET-33, LEU-65, GLN-66). Clearly, catalysis in LDH takes place in the regions where the fast vibrations of amino acids are preferentially concentrated.\\
%
%
\indent The structure of the localization pattern appears even more clearly when color-coding it onto the 3D~conformations (Fig.~\ref{fig:LDH-3DLL}), thus identifying unmistakably two distinct regions in the molecule where fast vibrations are concentrated. We observe here that peaks (hot spots) of the localization landscape that appear distant when plotted along the backbone chain (Fig.~\ref{fig:LDH-modes-LL}B) are found around the same spatial locations (here, the two red spots in Fig.~\ref{fig:LDH-3DLL}). We also find that the few peaks of the localization landscape that do not seem to correspond to any active site are in fact found in the same regions, once the backbone chain is folded into its tertiary structure. This observation applies very generally to all LLs computed for a very large set of enzymes (see Fig.~\ref{fig:partition-4enzymes} in the following).\\
%
%
\indent A careful analysis of the spatial structure of localized modes reveals that high-frequency localized vibrations are {\em compressive} motions. Hence, at hot-spots, amino acids tend to get close-packed. This feature is demonstrated here by computing at each residue the reduction of the mean distance between nearest neighbors induced by the highest-frequency modes (see Appendix~\ref{App:compression}). Figure \ref{fig:LDH-LL-compression} displays the result of this computation in the case of LDH: we clearly see that localization hot spots match almost exactly the regions subjected to compression motions of large magnitude. A more detailed analysis of a localized mode is presented in Fig.~\ref{fig:LDH-3D-compression} (the example shown in the figure is the eigenvector \#10).\\
%
\begin{figure}
\centering
\includegraphics[width=\columnwidth]{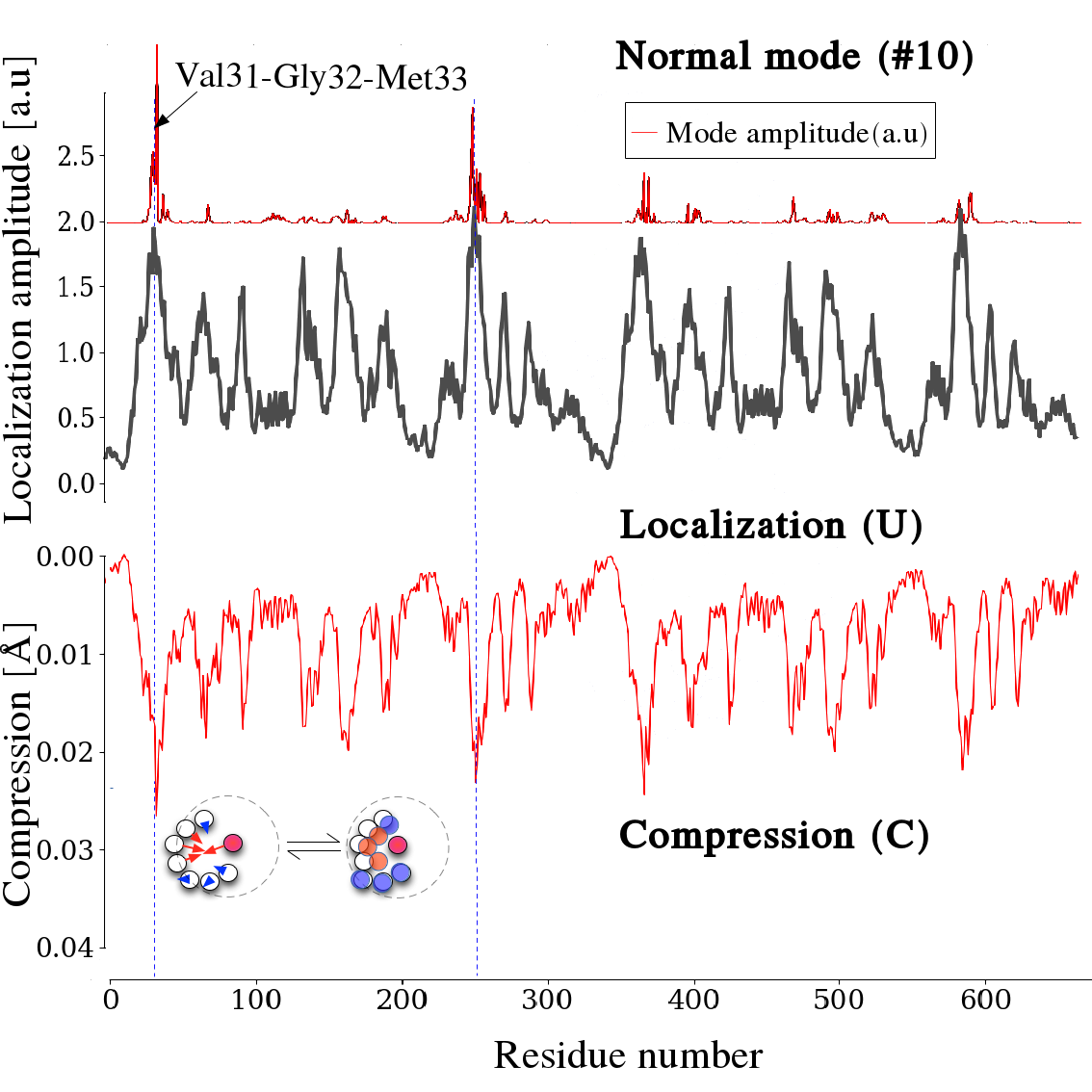}
\caption{\textbf{Compressive motions and localization sites in the L-Lactate Dehydrogenase dimer (LDH).} The displacement amplitude  (top graph) associated with the vibrational eigenvector \#10 (frequency \unit{94.17}{cm^{-1}}) is localized along the reaction coordinate residues VAL-31, GLY-32, MET-33, as predicted by the LL (function \textbf{U}, middle graph). The computation of the local compression factor (see Appendix~\ref{App:compression}) clearly shows that these localized modes are compression modes.}
\label{fig:LDH-LL-compression}
\end{figure}
%
%
\indent An important additional feature revealed by the LL analysis of LDH is that the enzyme structure appears to be partitioned into large-scale domains, i.e., contiguous sets of sites separated by deep minima of the landscape (Fig.~\ref{fig:LDH-domains}A). These domains comprise few hundreds of amino acids associated with the oligomeric complexes (monomer, dimer, trimer etc.). Each of these domains exhibits a sub-structure comprising 2 to 4 regions of a few tens of sites that harbor the most localized vibrations. From the LL, we can define each domain as comprising a hot-spot and extending to the two lowest local minima on both sides along the chain. Each of them can be understood as a nearly independent vibrational region (see Fig.~\ref{fig:LDH-domains}B), weakly coupled to its neighbors. This representation offers a totally new functional vision of the protein and also paves the way for a new understanding of allosteric processes~\cite{Yan2019}. This aspect will be addressed further in the Discussion section.\\
%
%
\indent The subtle connection unveiled above between localization of vibrational energy and compressive reorganization of the active site is by no means an isolated case. This has emerged neatly from the systematic study of a set of 933 enzymes from the catalytic site atlas~\cite{Porter2004}, comprising a total of 10,566 experimentally annotated catalytic sites. For each enzyme, we have computed the LL and located its highest maxima (examples of 3D representations of LLs for several enzymes are displayed in Fig.~\ref{fig:partition-4enzymes}, left column, while the right column displays the partitioning of each enzyme into independently vibrating domains, obtained from the LL using the procedure illustrated in Fig.~\ref{fig:LDH-domains}).\\
\indent Then, for each known catalytic site of the enzyme, we have computed the distance to the nearest maximum of the LL, expressed as a percent of the total length of the backbone chain (see Fig.~\ref{fig:proximity}A). Figure~\ref{fig:proximity}B displays a histogram of these relative distances, computed over all enzymes and all catalytic sites. The dotted curve plotted on top of the histogram represents the cumulative score. In 95\% of the cases, a catalytic site is found within 0.2\% of the total chain length from a localization hot spot. By comparison, the distance along the chain between a site picked at random and the nearest localized vibration site would be on average 10\% of the chain length, i.e., about 200~times farther away! This striking concordance clearly indicates that vibrational energy localization, as dictated by the 3D scaffold, must play a key role in the design of enzyme function: in 95\% of the case, catalytic sites are located in domains where residues exhibit fast compressive motions.
%
%

\section{Discussion}

\noindent Localization of vibrations is a general feature of the scaffold of proteins. The LL is a novel theoretical tool that allows one to capture quickly and efficiently the fundamental relationship between the 3D~structure and the spatial pattern of localized vibrations, first by predicting their locations and second by showing how the complex and irregular shape of the macromolecule can be partitioned (segmented) into a few weakly coupled clusters of vibrations. These are identified by highly localized vibrations involving few specific residues with periods of the order \unit{2-4}{ps}, that systematically take the form of compressive motions. Channeling thermal (or non-equilibrium) vibrational energy along such specific localized eigenvectors could be crucial for optimal enzyme functioning, e.g. in reducing the transfer distance associated with transition-state barriers or modulating donor-acceptor distances along specific directions, thus accelerating the chemical reaction step. Our analysis through the LL, performed on 933~enzymes, has confirmed  that the overwhelming majority of their catalytic sites are located at hot spots and are henceforth at the core of specific, fold-rooted compressive motions.\\
\indent These considerations can be given additional physical meaning in the context of a phenomenological modified Marcus-like tunneling theory that is used with success to interpret experimental data on enzyme-catalyzed H-transfer reactions~\cite{Meyer2005}. According to such theoretical scheme, the overall tunneling rate can be written as
\begin{equation}
\label{e:kgen}
k_t \propto e^{-\beta (\Delta G + \lambda)^2/4\lambda} \int e^{-S_G(R)/2\hbar}~\mathcal{P}_e(R) \, dR
\end{equation}
In the above expression $\Delta G$ denotes the free energy barrier associated with the global transition between reactant and product in the multi-dimensional space of heavy nuclear coordinate and $\lambda$ the corresponding reorganization energy, both associated with slow conformational sampling needed to reach the tunneling-ready state (TSR). The effect of rate-promoting vibrations is to weigh H tunneling from the ground-state, here expressed in the WKB approximation through the ground-state action $S_G(R)$ which is a function of the donor-acceptor distance $R$. The rate-promoting vibration(s) specifically couple to the DAD coordinate providing a slow modulation (compared to tunneling times) of the donor-acceptor potential energy represented by the equilibrium probability density $\mathcal{P}_e(R)$ corresponding to {\em optimal compression} through RPV motions  along the DAD at the active site.\\
\indent The characteristic times for thermally activated barrier crossing and/or tunneling in an enzymatic reaction are fast compared to the period of typical rate-promoting vibrations associated with the local reorganization of the active site (ps-ns),  which are themselves  swift compared to the time-scales of slow conformational sampling and conformational changes (ms-s). This hierarchy of time scales allows localized motions to slowly modulate (with respect to the actual transition step) the energy landscapes associated with chemical reactions. However, such modulations occur millions of times per second while the 3D conformation of the protein appears frozen, as the free energy landscape associated with the global reactant-product equilibrium is essentially static at the scale of the transition state lifetime. The striking and universal correspondence between the enzymatic active sites and the localization hot spots strongly suggests that such ps-ns local, time-modulated compressions are a basic feature of enzymes that is likely the product of evolutive optimization.\\
\indent Another intriguing logical consequence of our analysis is that resonance mechanisms (i.e the fact that clusters may eventually communicate with common vibrations) between distant localization sites may promote energy transfer across the molecular structure without affecting the sites located in between. By spatially confining vibrations at very specific places, wave localization may allow in principle distant sites to be ``fed'' with energy. Long-range communication would occur through specific protein paths associated with each specific frequency, without involving the rest of the structure (i.e thus preventing resonant leakage of energy to other modes). Therefore, localized vibrations may have a key role in allosteric effects, as pointed out in Ref.~\onlinecite{Piazza:2014aa}.\\
\indent In summary, investigating localized vibrations that control the active site reorganization in enzymes allows one to gain fundamental insight into the dynamical determinants of their functioning. The discovery of the related localization landscape sheds light onto the subtle link between the geography of fast compressive motions within an enzyme and its catalytic activity. Localized vibrations involving residues at or close to the active site correspond to motions that are typically compatible with the accepted timescales of rate-promoting vibrations (\unit{50-300}{cm^{-1}})~\cite{Schwartz2009, Klinman2013} and typically favor the shortening of transfer distances at molecular contact. Our analysis framework also offers an intriguing rationale for controlling fast dynamical effects at catalytic sites: any change in dynamical properties (interactions or mass) can be monitored with an extremely fast computational approach, allowing direct comparison with experiments, such as Kinetics Isotope Effects measurements~\cite{Nagel2009}.

\appendix

\section{Elastic network model of protein dynamics}
\label{App:ENM}

\noindent Elastic network models (ENM) of protein dynamics have been introduced by M. Tirion in 1996~\cite{Tirion1996} and later reformulated in a coarse-grained version by Bahar and co-workers under the name of anisotropic network model (ANM)~\cite{Bahar2005}. In the ANM, a given protein comprising $N$ residues is represented by an ensemble of $N$~fictitious particles, the mass of each particle being concentrated at the location of the corresponding $\alpha$-carbons. By definition, the equilibrium configuration of the system is taken to coincide with the experimentally solved structure (i.e. from X-ray diffraction or as an average over several NMR conformers). All particles are taken to have the same mass, which we set equal to the average amino acid mass $M = 110$ a.m.u., and each particle interacts with its neighboring particles through a central harmonic force. Let us denote $\rB_i(t)$ and $\RB_i$ the instantaneous and the equilibrium position vector of the $i$-th residue, respectively. The total potential energy of the system is that of a network of beads and central springs, that is,
\begin{equation}
\label{e:ANMtotpot}
V = \frac{1}{2} \sum_{i>j} K_{ij} (r_{ij} - R_{ij})^2 \,,
\end{equation}
where $K_{ij}$ is the force constant of the spring connecting the residues $i$ and $j$, while $r_{ij} = |\rB_i-\rB_j|$ and $R_{ij} = |\RB_i-\RB_j|$ are the instantaneous and equilibrium Euclidean distances between the pair $(i,j)$. The matrix of force constants can e specified in several ways. Here, in line with the original ideas of the ENM modeling strategy, we use a single stiffness~$k$ for all springs and identify the set of interacting pairs through a connectivity matrix, that is,
\begin{equation}
\label{e:ANMtotpot2}
K_{ij} = k \, c_{ij}
\end{equation}
where $c_{ij} = \{ 1 \ \text{for} \ R_{ij} \leq R_c \text{ and } 0 \ \text{otherwise} \}$. According to previous studies~\cite{Juanico2007}, we set $k=5$ kcal/mol/\AA$^2$ and choose a cutoff $R_c = 10$ \AA.
In order to compute the localization landscape of a protein, we consider the harmonic approximation of the ANM, which corresponds to 
\begin{equation}
\label{e:VtotTayl}
V = \frac{1}{2}\sum_{ij}\sum_{\alpha\beta}  \mathbb{H}_{ij}^{\alpha\beta} u_{i\alpha} u_{j\beta}
+ \mathcal{O}(u^{3})
\end{equation}
where $u_{i\alpha} = r_{i\alpha} - R_{i\alpha}$ ($\alpha=x,y,z$) are the Cartesian components of the displacement vector of residue $i$. The Hessian matrix $\mathbb{H}$ is directly derived from the total potential energy through
\begin{eqnarray}
\label{e:ANMHESS}
\mathbb{H}_{ij}^{\alpha\beta} 
&\stackrel{\rm def}{=}&
\left. \frac{\partial^{2} V}{\partial u_{i\alpha}\partial u_{j\beta}}\right|_{\{u=0\}} \nonumber\\
&=&  -K_{ij}  s^{\alpha}_{ij}s^{\beta}_{ij} + \delta_{ij} \sum_{m} K_{jm} s^{\alpha}_{mj} s^{\beta}_{mj}
\end{eqnarray}
where $s^{\alpha}_{ij} = R^{\alpha}_{ij}/R_{ij}$ are the Cartesian components of the unit equilibrium inter-particle vectors. The normal modes (NM) of a system of interacting particles, such as the residues in an elastic network, are the eigenvectors of the mass-weighted Hessian matrix (also known as dynamical matrix), 
\begin{equation}
\widetilde{\mathbb{H}} = M^{-1/2}~\mathbb{H}~M^{-1/2}  
\end{equation}
where $M$ is the diagonal mass matrix. It is well known that the high-frequency NMs of vibrations of protein structures are strongly localized in space, which is a result of the spatial quenched disorder of their equilibrium structures~\cite{Bahar2005}. This is still true in our coarse-grained model where the highest frequencies are of the order of \unit{100}{cm^{-1}} and the corresponding displacement vector fields are localized in regions of the size of one coordination shell, i.e. $\mathcal{O}(R_c)$. 

\section{The localization landscape of thermal phonons}
\label{App:LL}
 
\subsection{Calculation of the localization landscape}

\noindent Within the ANM framework, the equations of motion read
\begin{equation}\label{e:ANMDeqmot}
M_i \ddot u_{i\alpha} = -\sum_{j\beta} \mathbb{H}_{ij}^{\alpha\beta} u_{j\beta}
\end{equation}
By introducing the mass-weighted coordinates $X_{i\alpha} = \sqrt{M_i} u_{i\alpha}$,  this set of equations can be put into the following vector form:
\begin{equation}\label{eq:dynamic}
{\bf{\ddot X}} = -\widetilde{\mathbb{H}} \, \bf{X}
\end{equation}
We look for solutions to Eq.~(\ref{eq:dynamic})  in the form ${\bf X} = {\bf Y} e^{-j\omega t}$, which amounts to solving the related eigenvalue problem, i.e.  finding the eigenvectors ${\bf Y}^n$ and frequencies $\omega_n$ such that 
\begin{equation}\label{eq:EFDM2}
\widetilde{\mathbb{H}} \,{\bf Y}^n = \omega_n^2~{\bf Y}^n
\end{equation}
The displacement of residue $i$ can be decomposed into the contributions along each eigenvector ${\bf Y}^n$, that is,
\begin{equation}
u_{i\alpha}(t) = \frac{X_{i\alpha}(t)}{\sqrt{M_i}} = 
\frac{1}{\sqrt{M_i}} \sum\limits_{n=1}^{3N} \alpha_n Y^n_{i\alpha}~e^{-j\omega_n t}.
\end{equation}
Ref.~\cite{Filoche2012} introduces a mathematical function called localization landscape (LL) for predicting low-frequency localization. Yet, in the case of an inhomogeneous discrete system, high-frequency eigenvectors also correspond to localized, short-wavelength vibrations. According to a procedure similar to the one developed in~\cite{Lyra2015}, a high-frequency LL can also be computed as the solution~$\bf U$ to the following linear system
\begin{equation}
\label{eq:landscape}
\widetilde{\mathbb{H}}_c~{\bf U} = {\bf 1}\,,
\end{equation}
where
\begin{equation}
\widetilde{\mathbb{H}}^{\alpha\beta}_{c,i j}=
\begin{cases}
c-\widetilde{\mathbb{H}}^{\alpha\beta}_{i j} & \text{if } i=j,\alpha=\beta\\
\widetilde{\mathbb{H}}^{\alpha\beta}_{i j} & \text{otherwise}.
\end{cases}
\end{equation}
Here, $c$ is a small real positive constant such that all eigenvalues of the matrix $\widetilde{\mathbb{H}}_c$ are positive. 
The physical idea behind this (see Ref.~\onlinecite{Lyra2015}) is to look for localized modes of wave vector close 
to $k = \pi/a$ where $a \simeq$~\unit{3.83}{\AA} is the equilibrium distance between consecutive $\alpha$-carbons 
along the protein primary structure. This is the only 1D path belonging to the connectivity graph that ensures 
translational invariance along the chain. Finally, the localization landscape $\mathcal{U}$ used in this paper 
to rationalize the location of catalytic sites in enzymes is defined as the geometrical average of the three 
Cartesian components of ${\bf U}$, namely
\begin{equation}\label{e:scalarLL}
\mathcal{U}_i = \left( \sum_{\alpha\in x,y,z} U_{i\alpha}U_{i\alpha} \right)^{1/2}
\end{equation}

\section{Computing Efficiency of the Method}
\label{App:compeff}

\noindent An other important aspect of this approach is its remarkable computational efficiency. The study of proteins motions is usually conducted through an analysis of the normal modes. This requires solving the eigenvalue problem (see Eq.~\eqref{eq:EFDM2} in Appendix~\ref{App:LL})
\begin{equation}\label{eq:EFDM}
\widetilde{\mathbb{H}} \,{\bf Y}^n = \omega_n^2~{\bf Y}^n
\end{equation}
where ${\bf Y}$ and $\omega_n^2$ correspond to the normal modes and eigenfrequencies, respectively. Retrieving these quantities from normal modes analysis (NMA) can be a computational issue for large macromolecules (number of residues $N>10000$), especially when long range interactions are accounted for, as they considerably reduce the sparsity of the matrix~$\widetilde{\mathbb{H}}$. By contrast, the localization landscape is obtained by solving a simple linear system of algebraic equations
\begin{equation}
\hat L{\bf{U}} = {\bf{1}},
\end{equation}
where $\hat L$ stands for a self-adjoint operator constructed from the dynamical matrix (see Eq.~\eqref{eq:landscape} in Appendix~\ref{App:LL}). Table~\ref{tab:compare} compares the computational cost of the two aforementioned approaches, by reporting the required CPU-time as a function of the number of degrees of freedom (d.o.f). The ratio between the CPU times required by the two methods is displayed in the last column.
\begin{table}[htp]
\caption{\label{tab:compare} Comparison between 
Normal modes (NMA) and localization landscape (LL) analyses.}
\begin{ruledtabular}
\begin{tabular}{rrrc}
& \multicolumn{2}{c}{CPU time [s]} &  \\
\cmidrule{2-3}
$\#$ of d.o.f. &   NMA    &  LL                    &  Ratio NMA/LL \\ 
\cmidrule{1-4}
500              &  0.72    &  0.0032                &  22 \\
1000             &   4.6    &  0.17                  &  27 \\
2000             &    40    &     1                  &  40 \\
5000             &   840    &    18                  &  47 \\
10000            &  6600    &   132                  &  50 \\
20000            & 54000    &   571                  & 100 \\
\end{tabular}
\end{ruledtabular}
\end{table}
The LL approach is roughly 50 times more efficient for the typical protein size encountered in this study, although we have have restricted this analysis to the case of tridiagonal matrices: in practice, the computational gap between the two methods is even more substantial in realistic systems. This performance offers a clear advantage for a systematic analysis of large sets of protein data.

%
\begin{figure}[t!]
\centering
\includegraphics[width=\columnwidth]{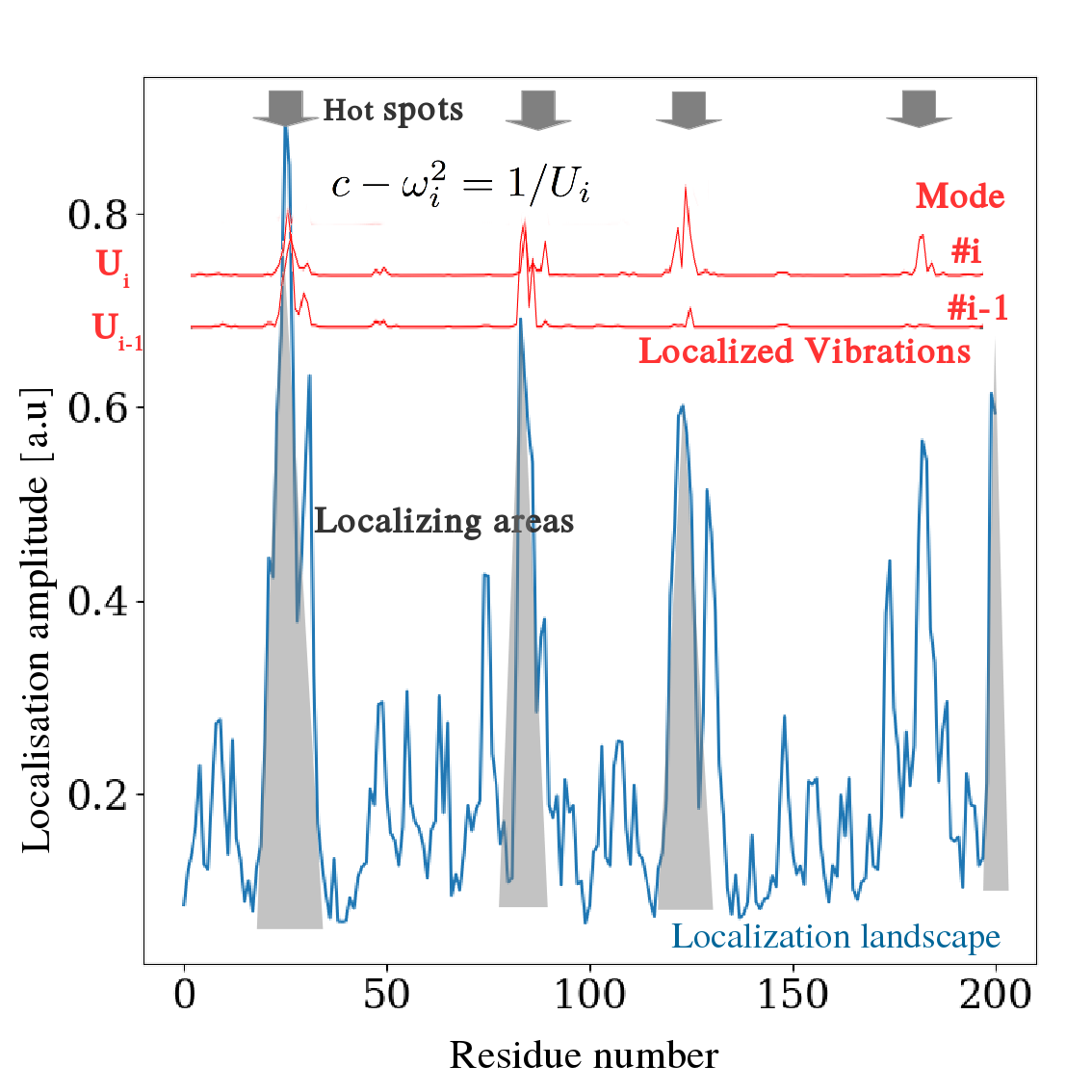}
\caption{\textbf{Localization Landscape for HIV-1 Protease (PDB id: 1A30)}. Wave localization is visualized through the displacement pattern of the 2~fastest eigenmodes (1 and 2), with frequencies of the order of \unit{96}{cm^{-1}}. 
The locations of the maxima identify the most localizing areas, i.e. the ``hot spots''. Each eigenfrequency can be 
associated with a peak height in the landscape, whose values are arbitrary as they depend on the choice 
of the constant $c$ in Eq.~\eqref{eq:landscape}. In this case, we have chosen $c=18$ 
in non-dimensional units ($k=M=1$), to ensure that all eigenvalues of the operator~\eqref{eq:landscape} are positive.
The maximum value of $U$ ($\approx 0.9$) yields $\omega_{\rm max} = \sqrt{c - 1/U_{\rm max}}\approx 4.11$. 
With the choice $k = 5$ kcal/mol/\AA$^2$, $M=110$ a.m.u., this gives $\omega_{\rm max} \approx$ \unit{94.8}{cm^{-1}}, 
in agreement with the maximum frequency found by brute-force diagonalization of the dynamical matrix.  
The line that cuts horizontally the landscape at a given height reveals where the vibrations 
at that particular frequency are observed along the backbone chain.}
\label{fig:annex}
\end{figure}

\section{Calculation of the local compression factor}
\label{App:compression}

\noindent The compression factor $\mathcal{C}_i$ measures the average level of local compression at a given site. For a given pair $i,j$, this amounts to evaluating the change in Euclidean distance along a given normal mode with respect to the equilibrium distance $R_{ij}$. In mathematical terms, $C_i$ reads
\begin{widetext}
\begin{equation}\label{e:scalarLL2}
\mathcal{C}_i = \frac{1}{N_{\mathcal{S}} \,c_i} \sum_{n\in\mathcal{S}} \sum_{j} c_{ij} 
\left[ R_{ij}  - 
    \left( 
        \sum_{\alpha=x,y,z}
             (R^\alpha_{ij} + a({Y}^n_{i\alpha} - {Y}^n_{j\alpha}))^2
    \right)^{1/2}
\right]\,,
\end{equation}       
\end{widetext}
where $\mathcal{S}$ is the set comprising the $N_{\mathcal{S}}$ highest-frequency normal modes, $c_{i} = \sum_j c_{ij}$ is the connectivity of residue $i$ and $a$ is an arbitrary displacement in \AA. In our calculation we chose $a = 1$ \AA, smaller than half the shortest inter-residue distance $R_{ij} \simeq 3.8$ \AA. This ensures that $\mathcal{C}_i$ are positive quantities, in agreement with the physical requirement that relative displacements cannot exceed equilibrium inter-distances.

\begin{acknowledgments}
S. M. is funded by a NSF INSPIRE grant and a Simons fellowship. S. M., C.W., and M. F. are funded by a grant from the Simons Foundation (563916, SM, 601954, CW, and 601944, MF).
\end{acknowledgments}


\begin{thebibliography}{52}%
\makeatletter
\providecommand \@ifxundefined [1]{%
 \@ifx{#1\undefined}
}%
\providecommand \@ifnum [1]{%
 \ifnum #1\expandafter \@firstoftwo
 \else \expandafter \@secondoftwo
 \fi
}%
\providecommand \@ifx [1]{%
 \ifx #1\expandafter \@firstoftwo
 \else \expandafter \@secondoftwo
 \fi
}%
\providecommand \natexlab [1]{#1}%
\providecommand \enquote  [1]{``#1''}%
\providecommand \bibnamefont  [1]{#1}%
\providecommand \bibfnamefont [1]{#1}%
\providecommand \citenamefont [1]{#1}%
\providecommand \href@noop [0]{\@secondoftwo}%
\providecommand \href [0]{\begingroup \@sanitize@url \@href}%
\providecommand \@href[1]{\@@startlink{#1}\@@href}%
\providecommand \@@href[1]{\endgroup#1\@@endlink}%
\providecommand \@sanitize@url [0]{\catcode `\\12\catcode `\$12\catcode
  `\&12\catcode `\#12\catcode `\^12\catcode `\_12\catcode `\%12\relax}%
\providecommand \@@startlink[1]{}%
\providecommand \@@endlink[0]{}%
\providecommand \url  [0]{\begingroup\@sanitize@url \@url }%
\providecommand \@url [1]{\endgroup\@href {#1}{\urlprefix }}%
\providecommand \urlprefix  [0]{URL }%
\providecommand \Eprint [0]{\href }%
\providecommand \doibase [0]{http://dx.doi.org/}%
\providecommand \selectlanguage [0]{\@gobble}%
\providecommand \bibinfo  [0]{\@secondoftwo}%
\providecommand \bibfield  [0]{\@secondoftwo}%
\providecommand \translation [1]{[#1]}%
\providecommand \BibitemOpen [0]{}%
\providecommand \bibitemStop [0]{}%
\providecommand \bibitemNoStop [0]{.\EOS\space}%
\providecommand \EOS [0]{\spacefactor3000\relax}%
\providecommand \BibitemShut  [1]{\csname bibitem#1\endcsname}%
\let\auto@bib@innerbib\@empty
\bibitem [{\citenamefont {Nagel}\ and\ \citenamefont
  {Klinman}(2009)}]{Nagel2009}%
  \BibitemOpen
  \bibfield  {author} {\bibinfo {author} {\bibfnamefont {Z.~D.}\ \bibnamefont
  {Nagel}}\ and\ \bibinfo {author} {\bibfnamefont {J.~P.}\ \bibnamefont
  {Klinman}},\ }\href@noop {} {\bibfield  {journal} {\bibinfo  {journal}
  {Nature Chemical Biology}\ }\textbf {\bibinfo {volume} {5}},\ \bibinfo
  {pages} {543} (\bibinfo {year} {2009})}\BibitemShut {NoStop}%
\bibitem [{\citenamefont {Oldfield}\ and\ \citenamefont
  {Dunker}(2014)}]{Oldfield2014}%
  \BibitemOpen
  \bibfield  {author} {\bibinfo {author} {\bibfnamefont {C.~J.}\ \bibnamefont
  {Oldfield}}\ and\ \bibinfo {author} {\bibfnamefont {A.~K.}\ \bibnamefont
  {Dunker}},\ }\href@noop {} {\bibfield  {journal} {\bibinfo  {journal} {Annual
  Review of Biochemistry}\ }\textbf {\bibinfo {volume} {83}},\ \bibinfo {pages}
  {553} (\bibinfo {year} {2014})}\BibitemShut {NoStop}%
\bibitem [{\citenamefont {Zinovjev}\ and\ \citenamefont
  {Tu{\~n}{\'o}n}(2017)}]{Zinovjev2017}%
  \BibitemOpen
  \bibfield  {author} {\bibinfo {author} {\bibfnamefont {K.}~\bibnamefont
  {Zinovjev}}\ and\ \bibinfo {author} {\bibfnamefont {I.}~\bibnamefont
  {Tu{\~n}{\'o}n}},\ }\href@noop {} {\bibfield  {journal} {\bibinfo  {journal}
  {Proceedings of the National Academy of Sciences}\ }\textbf {\bibinfo
  {volume} {114}},\ \bibinfo {pages} {12390} (\bibinfo {year}
  {2017})}\BibitemShut {NoStop}%
\bibitem [{\citenamefont {Kale}\ \emph {et~al.}(2008)\citenamefont {Kale},
  \citenamefont {Ulas}, \citenamefont {Song}, \citenamefont {Brudvig},
  \citenamefont {Furey},\ and\ \citenamefont {Jordan}}]{Kale2008}%
  \BibitemOpen
  \bibfield  {author} {\bibinfo {author} {\bibfnamefont {S.}~\bibnamefont
  {Kale}}, \bibinfo {author} {\bibfnamefont {G.}~\bibnamefont {Ulas}}, \bibinfo
  {author} {\bibfnamefont {J.}~\bibnamefont {Song}}, \bibinfo {author}
  {\bibfnamefont {G.~W.}\ \bibnamefont {Brudvig}}, \bibinfo {author}
  {\bibfnamefont {W.}~\bibnamefont {Furey}}, \ and\ \bibinfo {author}
  {\bibfnamefont {F.}~\bibnamefont {Jordan}},\ }\href@noop {} {\bibfield
  {journal} {\bibinfo  {journal} {Proceedings of the National Academy of
  Sciences}\ }\textbf {\bibinfo {volume} {105}},\ \bibinfo {pages} {1158}
  (\bibinfo {year} {2008})}\BibitemShut {NoStop}%
\bibitem [{\citenamefont {Agarwal}(2005)}]{Agarwal2005}%
  \BibitemOpen
  \bibfield  {author} {\bibinfo {author} {\bibfnamefont {P.~K.}\ \bibnamefont
  {Agarwal}},\ }\href@noop {} {\bibfield  {journal} {\bibinfo  {journal}
  {Journal of the American Chemical Society}\ }\textbf {\bibinfo {volume}
  {127}},\ \bibinfo {pages} {15248} (\bibinfo {year} {2005})}\BibitemShut
  {NoStop}%
\bibitem [{\citenamefont {Antoniou}\ and\ \citenamefont
  {Schwartz}(2001)}]{Antoniou2001}%
  \BibitemOpen
  \bibfield  {author} {\bibinfo {author} {\bibfnamefont {D.}~\bibnamefont
  {Antoniou}}\ and\ \bibinfo {author} {\bibfnamefont {S.~D.}\ \bibnamefont
  {Schwartz}},\ }\href@noop {} {\bibfield  {journal} {\bibinfo  {journal} {The
  Journal of Physical Chemistry B}\ }\textbf {\bibinfo {volume} {105}},\
  \bibinfo {pages} {5553} (\bibinfo {year} {2001})}\BibitemShut {NoStop}%
\bibitem [{\citenamefont {Hay}\ and\ \citenamefont {Scrutton}(2012)}]{Hay2012}%
  \BibitemOpen
  \bibfield  {author} {\bibinfo {author} {\bibfnamefont {S.}~\bibnamefont
  {Hay}}\ and\ \bibinfo {author} {\bibfnamefont {N.~S.}\ \bibnamefont
  {Scrutton}},\ }\href@noop {} {\bibfield  {journal} {\bibinfo  {journal}
  {Nature Chemistry}\ }\textbf {\bibinfo {volume} {4}},\ \bibinfo {pages} {161}
  (\bibinfo {year} {2012})}\BibitemShut {NoStop}%
\bibitem [{\citenamefont {Luk}\ \emph {et~al.}(2013)\citenamefont {Luk},
  \citenamefont {Javier Ruiz-Pern{\'\i}a}, \citenamefont {Dawson},
  \citenamefont {Roca}, \citenamefont {Loveridge}, \citenamefont {Glowacki},
  \citenamefont {Harvey}, \citenamefont {Mulholland}, \citenamefont
  {Tu{\~n}{\'o}n}, \citenamefont {Moliner},\ and\ \citenamefont
  {Allemann}}]{Luk2013}%
  \BibitemOpen
  \bibfield  {author} {\bibinfo {author} {\bibfnamefont {L.~Y.~P.}\
  \bibnamefont {Luk}}, \bibinfo {author} {\bibfnamefont {J.}~\bibnamefont
  {Javier Ruiz-Pern{\'\i}a}}, \bibinfo {author} {\bibfnamefont {W.~M.}\
  \bibnamefont {Dawson}}, \bibinfo {author} {\bibfnamefont {M.}~\bibnamefont
  {Roca}}, \bibinfo {author} {\bibfnamefont {E.~J.}\ \bibnamefont {Loveridge}},
  \bibinfo {author} {\bibfnamefont {D.~R.}\ \bibnamefont {Glowacki}}, \bibinfo
  {author} {\bibfnamefont {J.~N.}\ \bibnamefont {Harvey}}, \bibinfo {author}
  {\bibfnamefont {A.~J.}\ \bibnamefont {Mulholland}}, \bibinfo {author}
  {\bibfnamefont {I.}~\bibnamefont {Tu{\~n}{\'o}n}}, \bibinfo {author}
  {\bibfnamefont {V.}~\bibnamefont {Moliner}}, \ and\ \bibinfo {author}
  {\bibfnamefont {R.~K.}\ \bibnamefont {Allemann}},\ }\href@noop {} {\bibfield
  {journal} {\bibinfo  {journal} {Proceedings of the National Academy of
  Sciences}\ }\textbf {\bibinfo {volume} {110}},\ \bibinfo {pages} {16344}
  (\bibinfo {year} {2013})}\BibitemShut {NoStop}%
\bibitem [{\citenamefont {Pudney}\ \emph {et~al.}(2009)\citenamefont {Pudney},
  \citenamefont {Hay}, \citenamefont {Levy}, \citenamefont {Pang},
  \citenamefont {Sutcliffe}, \citenamefont {Leys},\ and\ \citenamefont
  {Scrutton}}]{Pudney2009}%
  \BibitemOpen
  \bibfield  {author} {\bibinfo {author} {\bibfnamefont {C.~R.}\ \bibnamefont
  {Pudney}}, \bibinfo {author} {\bibfnamefont {S.}~\bibnamefont {Hay}},
  \bibinfo {author} {\bibfnamefont {C.}~\bibnamefont {Levy}}, \bibinfo {author}
  {\bibfnamefont {J.}~\bibnamefont {Pang}}, \bibinfo {author} {\bibfnamefont
  {M.~J.}\ \bibnamefont {Sutcliffe}}, \bibinfo {author} {\bibfnamefont
  {D.}~\bibnamefont {Leys}}, \ and\ \bibinfo {author} {\bibfnamefont {N.~S.}\
  \bibnamefont {Scrutton}},\ }\href@noop {} {\bibfield  {journal} {\bibinfo
  {journal} {Journal of the American Chemical Society}\ }\textbf {\bibinfo
  {volume} {131}},\ \bibinfo {pages} {17072} (\bibinfo {year}
  {2009})}\BibitemShut {NoStop}%
\bibitem [{\citenamefont {Heyes}\ \emph {et~al.}(2009)\citenamefont {Heyes},
  \citenamefont {Sakuma}, \citenamefont {de~Visser},\ and\ \citenamefont
  {Scrutton}}]{Heyes2009}%
  \BibitemOpen
  \bibfield  {author} {\bibinfo {author} {\bibfnamefont {D.~J.}\ \bibnamefont
  {Heyes}}, \bibinfo {author} {\bibfnamefont {M.}~\bibnamefont {Sakuma}},
  \bibinfo {author} {\bibfnamefont {S.~P.}\ \bibnamefont {de~Visser}}, \ and\
  \bibinfo {author} {\bibfnamefont {N.~S.}\ \bibnamefont {Scrutton}},\
  }\href@noop {} {\bibfield  {journal} {\bibinfo  {journal} {Journal of
  Biological Chemistry}\ }\textbf {\bibinfo {volume} {284}},\ \bibinfo {pages}
  {3762} (\bibinfo {year} {2009})}\BibitemShut {NoStop}%
\bibitem [{\citenamefont {Heyes}\ \emph {et~al.}(2011)\citenamefont {Heyes},
  \citenamefont {Levy}, \citenamefont {Sakuma}, \citenamefont {Robertson},\
  and\ \citenamefont {Scrutton}}]{Heyes2011}%
  \BibitemOpen
  \bibfield  {author} {\bibinfo {author} {\bibfnamefont {D.~J.}\ \bibnamefont
  {Heyes}}, \bibinfo {author} {\bibfnamefont {C.}~\bibnamefont {Levy}},
  \bibinfo {author} {\bibfnamefont {M.}~\bibnamefont {Sakuma}}, \bibinfo
  {author} {\bibfnamefont {D.~L.}\ \bibnamefont {Robertson}}, \ and\ \bibinfo
  {author} {\bibfnamefont {N.~S.}\ \bibnamefont {Scrutton}},\ }\href@noop {}
  {\bibfield  {journal} {\bibinfo  {journal} {Journal of Biological Chemistry}\
  }\textbf {\bibinfo {volume} {286}},\ \bibinfo {pages} {11849} (\bibinfo
  {year} {2011})}\BibitemShut {NoStop}%
\bibitem [{\citenamefont {Henzler-Wildman}\ \emph {et~al.}(2018)\citenamefont
  {Henzler-Wildman}, \citenamefont {Thai}, \citenamefont {Lei}, \citenamefont
  {Ott}, \citenamefont {Wolf-Watz}, \citenamefont {Fenn}, \citenamefont
  {Pozharski}, \citenamefont {Wilson}, \citenamefont {Petsko}, \citenamefont
  {Karplus}, \citenamefont {H\"ubner},\ and\ \citenamefont
  {Kern}}]{Henzler-Wildman2007c}%
  \BibitemOpen
  \bibfield  {author} {\bibinfo {author} {\bibfnamefont {K.~A.}\ \bibnamefont
  {Henzler-Wildman}}, \bibinfo {author} {\bibfnamefont {V.}~\bibnamefont
  {Thai}}, \bibinfo {author} {\bibfnamefont {M.}~\bibnamefont {Lei}}, \bibinfo
  {author} {\bibfnamefont {M.}~\bibnamefont {Ott}}, \bibinfo {author}
  {\bibfnamefont {M.}~\bibnamefont {Wolf-Watz}}, \bibinfo {author}
  {\bibfnamefont {T.}~\bibnamefont {Fenn}}, \bibinfo {author} {\bibfnamefont
  {E.}~\bibnamefont {Pozharski}}, \bibinfo {author} {\bibfnamefont {M.~A.}\
  \bibnamefont {Wilson}}, \bibinfo {author} {\bibfnamefont {G.~A.}\
  \bibnamefont {Petsko}}, \bibinfo {author} {\bibfnamefont {M.}~\bibnamefont
  {Karplus}}, \bibinfo {author} {\bibfnamefont {C.~G.}\ \bibnamefont
  {H\"ubner}}, \ and\ \bibinfo {author} {\bibfnamefont {D.}~\bibnamefont
  {Kern}},\ }\href@noop {} {\bibfield  {journal} {\bibinfo  {journal} {Nature}\
  }\textbf {\bibinfo {volume} {450}},\ \bibinfo {pages} {838} (\bibinfo {year}
  {2018})}\BibitemShut {NoStop}%
\bibitem [{\citenamefont {Saen-Oon}\ \emph {et~al.}(2008)\citenamefont
  {Saen-Oon}, \citenamefont {Ghanem}, \citenamefont {Schramm},\ and\
  \citenamefont {Schwartz}}]{Saen-Oon2008}%
  \BibitemOpen
  \bibfield  {author} {\bibinfo {author} {\bibfnamefont {S.}~\bibnamefont
  {Saen-Oon}}, \bibinfo {author} {\bibfnamefont {M.}~\bibnamefont {Ghanem}},
  \bibinfo {author} {\bibfnamefont {V.~L.}\ \bibnamefont {Schramm}}, \ and\
  \bibinfo {author} {\bibfnamefont {S.~D.}\ \bibnamefont {Schwartz}},\ }\href
  {\doibase 10.1529/biophysj.107.121913} {\bibfield  {journal} {\bibinfo
  {journal} {Biophysical Journal}\ }\textbf {\bibinfo {volume} {94}},\ \bibinfo
  {pages} {4078} (\bibinfo {year} {2008})}\BibitemShut {NoStop}%
\bibitem [{\citenamefont {Masterson}\ \emph {et~al.}(2010)\citenamefont
  {Masterson}, \citenamefont {Cheng}, \citenamefont {Yu}, \citenamefont
  {Tonelli}, \citenamefont {Kornev}, \citenamefont {Taylor},\ and\
  \citenamefont {Veglia}}]{Masterson2010}%
  \BibitemOpen
  \bibfield  {author} {\bibinfo {author} {\bibfnamefont {L.~R.}\ \bibnamefont
  {Masterson}}, \bibinfo {author} {\bibfnamefont {C.}~\bibnamefont {Cheng}},
  \bibinfo {author} {\bibfnamefont {T.}~\bibnamefont {Yu}}, \bibinfo {author}
  {\bibfnamefont {M.}~\bibnamefont {Tonelli}}, \bibinfo {author} {\bibfnamefont
  {A.}~\bibnamefont {Kornev}}, \bibinfo {author} {\bibfnamefont {S.~S.}\
  \bibnamefont {Taylor}}, \ and\ \bibinfo {author} {\bibfnamefont
  {G.}~\bibnamefont {Veglia}},\ }\href@noop {} {\bibfield  {journal} {\bibinfo
  {journal} {Nature Chemical Biology}\ }\textbf {\bibinfo {volume} {6}},\
  \bibinfo {pages} {821} (\bibinfo {year} {2010})}\BibitemShut {NoStop}%
\bibitem [{\citenamefont {Agarwal}\ \emph {et~al.}(2002)\citenamefont
  {Agarwal}, \citenamefont {Billeter}, \citenamefont {Rajagopalan},
  \citenamefont {Benkovic},\ and\ \citenamefont
  {Hammes-Schiffer}}]{Agarwal2002}%
  \BibitemOpen
  \bibfield  {author} {\bibinfo {author} {\bibfnamefont {P.~K.}\ \bibnamefont
  {Agarwal}}, \bibinfo {author} {\bibfnamefont {S.~R.}\ \bibnamefont
  {Billeter}}, \bibinfo {author} {\bibfnamefont {P.~T.~R.}\ \bibnamefont
  {Rajagopalan}}, \bibinfo {author} {\bibfnamefont {S.~J.}\ \bibnamefont
  {Benkovic}}, \ and\ \bibinfo {author} {\bibfnamefont {S.}~\bibnamefont
  {Hammes-Schiffer}},\ }\href@noop {} {\bibfield  {journal} {\bibinfo
  {journal} {Proceedings of the National Academy of Sciences}\ }\textbf
  {\bibinfo {volume} {99}},\ \bibinfo {pages} {2794} (\bibinfo {year}
  {2002})}\BibitemShut {NoStop}%
\bibitem [{\citenamefont {McClare}(1972)}]{McClare1972}%
  \BibitemOpen
  \bibfield  {author} {\bibinfo {author} {\bibfnamefont {C.}~\bibnamefont
  {McClare}},\ }\href {\doibase https://doi.org/10.1016/0022-5193(72)90151-8}
  {\bibfield  {journal} {\bibinfo  {journal} {Journal of Theoretical Biology}\
  }\textbf {\bibinfo {volume} {35}},\ \bibinfo {pages} {569 } (\bibinfo {year}
  {1972})}\BibitemShut {NoStop}%
\bibitem [{\citenamefont {Chen}\ and\ \citenamefont
  {Schwartz}(2018)}]{Chen2018}%
  \BibitemOpen
  \bibfield  {author} {\bibinfo {author} {\bibfnamefont {X.}~\bibnamefont
  {Chen}}\ and\ \bibinfo {author} {\bibfnamefont {S.~D.}\ \bibnamefont
  {Schwartz}},\ }\href@noop {} {\bibfield  {journal} {\bibinfo  {journal}
  {Biochemistry}\ }\textbf {\bibinfo {volume} {57}},\ \bibinfo {pages} {3289}
  (\bibinfo {year} {2018})}\BibitemShut {NoStop}%
\bibitem [{\citenamefont {Dzierlenga}\ and\ \citenamefont
  {Schwartz}(2016)}]{Dzierlenga2016}%
  \BibitemOpen
  \bibfield  {author} {\bibinfo {author} {\bibfnamefont {M.~W.}\ \bibnamefont
  {Dzierlenga}}\ and\ \bibinfo {author} {\bibfnamefont {S.~D.}\ \bibnamefont
  {Schwartz}},\ }\href@noop {} {\bibfield  {journal} {\bibinfo  {journal} {The
  Journal of Physical Chemistry Letters}\ }\textbf {\bibinfo {volume} {7}},\
  \bibinfo {pages} {2591} (\bibinfo {year} {2016})}\BibitemShut {NoStop}%
\bibitem [{\citenamefont {Quaytman}\ and\ \citenamefont
  {Schwartz}(2007)}]{Quaytman2007}%
  \BibitemOpen
  \bibfield  {author} {\bibinfo {author} {\bibfnamefont {S.~L.}\ \bibnamefont
  {Quaytman}}\ and\ \bibinfo {author} {\bibfnamefont {S.~D.}\ \bibnamefont
  {Schwartz}},\ }\href@noop {} {\bibfield  {journal} {\bibinfo  {journal}
  {Proceedings of the National Academy of Sciences}\ }\textbf {\bibinfo
  {volume} {104}},\ \bibinfo {pages} {12253} (\bibinfo {year}
  {2007})}\BibitemShut {NoStop}%
\bibitem [{\citenamefont {Harijan}\ \emph {et~al.}(2017)\citenamefont
  {Harijan}, \citenamefont {Zoi}, \citenamefont {Antoniou}, \citenamefont
  {Schwartz},\ and\ \citenamefont {Schramm}}]{Harijan2017}%
  \BibitemOpen
  \bibfield  {author} {\bibinfo {author} {\bibfnamefont {R.~K.}\ \bibnamefont
  {Harijan}}, \bibinfo {author} {\bibfnamefont {I.}~\bibnamefont {Zoi}},
  \bibinfo {author} {\bibfnamefont {D.}~\bibnamefont {Antoniou}}, \bibinfo
  {author} {\bibfnamefont {S.~D.}\ \bibnamefont {Schwartz}}, \ and\ \bibinfo
  {author} {\bibfnamefont {V.~L.}\ \bibnamefont {Schramm}},\ }\href@noop {}
  {\bibfield  {journal} {\bibinfo  {journal} {Proceedings of the National
  Academy of Sciences}\ }\textbf {\bibinfo {volume} {114}},\ \bibinfo {pages}
  {6456} (\bibinfo {year} {2017})}\BibitemShut {NoStop}%
\bibitem [{\citenamefont {Arcus}\ and\ \citenamefont
  {Pudney}(2015)}]{Arcus2015}%
  \BibitemOpen
  \bibfield  {author} {\bibinfo {author} {\bibfnamefont {V.~L.}\ \bibnamefont
  {Arcus}}\ and\ \bibinfo {author} {\bibfnamefont {C.~R.}\ \bibnamefont
  {Pudney}},\ }\href@noop {} {\bibfield  {journal} {\bibinfo  {journal} {FEBS
  Letters}\ }\textbf {\bibinfo {volume} {589}},\ \bibinfo {pages} {2200}
  (\bibinfo {year} {2015})}\BibitemShut {NoStop}%
\bibitem [{\citenamefont {Bruno}\ and\ \citenamefont
  {Bialek}(1992)}]{Bruno1992}%
  \BibitemOpen
  \bibfield  {author} {\bibinfo {author} {\bibfnamefont {W.~J.}\ \bibnamefont
  {Bruno}}\ and\ \bibinfo {author} {\bibfnamefont {W.}~\bibnamefont {Bialek}},\
  }\href@noop {} {\bibfield  {journal} {\bibinfo  {journal} {Biophysical
  Journal}\ }\textbf {\bibinfo {volume} {63}},\ \bibinfo {pages} {689}
  (\bibinfo {year} {1992})}\BibitemShut {NoStop}%
\bibitem [{\citenamefont {McCammon}\ and\ \citenamefont
  {Harvey}(1987)}]{McCammon:1987aa}%
  \BibitemOpen
  \bibfield  {author} {\bibinfo {author} {\bibfnamefont {J.~A.}\ \bibnamefont
  {McCammon}}\ and\ \bibinfo {author} {\bibfnamefont {S.~C.}\ \bibnamefont
  {Harvey}},\ }\href@noop {} {\emph {\bibinfo {title} {Dynamics of Proteins and
  Nucleic Acids}}}\ (\bibinfo  {publisher} {Cambridge University Press},\
  \bibinfo {address} {New York},\ \bibinfo {year} {1987})\BibitemShut {NoStop}%
\bibitem [{\citenamefont {Wolf-Watz}\ \emph {et~al.}(2004)\citenamefont
  {Wolf-Watz}, \citenamefont {Thai}, \citenamefont {Henzler-Wildman},
  \citenamefont {Hadjipavlou}, \citenamefont {Eisenmesser},\ and\ \citenamefont
  {Kern}}]{Wolf-Watz2004}%
  \BibitemOpen
  \bibfield  {author} {\bibinfo {author} {\bibfnamefont {M.}~\bibnamefont
  {Wolf-Watz}}, \bibinfo {author} {\bibfnamefont {V.}~\bibnamefont {Thai}},
  \bibinfo {author} {\bibfnamefont {K.}~\bibnamefont {Henzler-Wildman}},
  \bibinfo {author} {\bibfnamefont {G.}~\bibnamefont {Hadjipavlou}}, \bibinfo
  {author} {\bibfnamefont {E.~Z.}\ \bibnamefont {Eisenmesser}}, \ and\ \bibinfo
  {author} {\bibfnamefont {D.}~\bibnamefont {Kern}},\ }\href@noop {} {\bibfield
   {journal} {\bibinfo  {journal} {Nature Structural and Molecular Biology}\
  }\textbf {\bibinfo {volume} {11}},\ \bibinfo {pages} {945} (\bibinfo {year}
  {2004})}\BibitemShut {NoStop}%
\bibitem [{\citenamefont {Changeux}\ and\ \citenamefont
  {Edelstein}(2005)}]{Changeux2005}%
  \BibitemOpen
  \bibfield  {author} {\bibinfo {author} {\bibfnamefont {J.-P.}\ \bibnamefont
  {Changeux}}\ and\ \bibinfo {author} {\bibfnamefont {S.~J.}\ \bibnamefont
  {Edelstein}},\ }\href {\doibase 10.1126/science.1108595} {\bibfield
  {journal} {\bibinfo  {journal} {Science (New York, N.Y.)}\ }\textbf {\bibinfo
  {volume} {308}},\ \bibinfo {pages} {1424} (\bibinfo {year}
  {2005})}\BibitemShut {NoStop}%
\bibitem [{\citenamefont {Hammes}(2002)}]{Hammes2002}%
  \BibitemOpen
  \bibfield  {author} {\bibinfo {author} {\bibfnamefont {G.~G.}\ \bibnamefont
  {Hammes}},\ }\href@noop {} {\bibfield  {journal} {\bibinfo  {journal}
  {Biochemistry}\ }\textbf {\bibinfo {volume} {41}},\ \bibinfo {pages} {8221}
  (\bibinfo {year} {2002})}\BibitemShut {NoStop}%
\bibitem [{\citenamefont {Gerhart}\ and\ \citenamefont
  {Schachman}(1968)}]{Gerhart1968}%
  \BibitemOpen
  \bibfield  {author} {\bibinfo {author} {\bibfnamefont {J.~C.}\ \bibnamefont
  {Gerhart}}\ and\ \bibinfo {author} {\bibfnamefont {H.~K.}\ \bibnamefont
  {Schachman}},\ }\href@noop {} {\bibfield  {journal} {\bibinfo  {journal}
  {Biochemistry}\ }\textbf {\bibinfo {volume} {7}},\ \bibinfo {pages} {538}
  (\bibinfo {year} {1968})}\BibitemShut {NoStop}%
\bibitem [{\citenamefont {English}\ \emph {et~al.}(2006)\citenamefont
  {English}, \citenamefont {Min}, \citenamefont {{Van Oijen}}, \citenamefont
  {Kang}, \citenamefont {Luo}, \citenamefont {Sun}, \citenamefont {Cherayil},
  \citenamefont {Kou},\ and\ \citenamefont {Xie}}]{English2006}%
  \BibitemOpen
  \bibfield  {author} {\bibinfo {author} {\bibfnamefont {B.~P.}\ \bibnamefont
  {English}}, \bibinfo {author} {\bibfnamefont {W.}~\bibnamefont {Min}},
  \bibinfo {author} {\bibfnamefont {A.~M.}\ \bibnamefont {{Van Oijen}}},
  \bibinfo {author} {\bibfnamefont {T.~L.}\ \bibnamefont {Kang}}, \bibinfo
  {author} {\bibfnamefont {G.}~\bibnamefont {Luo}}, \bibinfo {author}
  {\bibfnamefont {H.}~\bibnamefont {Sun}}, \bibinfo {author} {\bibfnamefont
  {B.~J.}\ \bibnamefont {Cherayil}}, \bibinfo {author} {\bibfnamefont {S.~C.}\
  \bibnamefont {Kou}}, \ and\ \bibinfo {author} {\bibfnamefont {X.~S.}\
  \bibnamefont {Xie}},\ }\href@noop {} {\bibfield  {journal} {\bibinfo
  {journal} {Nature Chemical Biology}\ }\textbf {\bibinfo {volume} {2}},\
  \bibinfo {pages} {87} (\bibinfo {year} {2006})}\BibitemShut {NoStop}%
\bibitem [{\citenamefont {{Lu, H. P. Luying Xun}}(1998)}]{Xun:1998aa}%
  \BibitemOpen
  \bibfield  {author} {\bibinfo {author} {\bibfnamefont {X.~S.~X.}\
  \bibnamefont {{Lu, H. P. Luying Xun}}},\ }\href@noop {} {\bibfield  {journal}
  {\bibinfo  {journal} {Science}\ }\textbf {\bibinfo {volume} {282}},\ \bibinfo
  {pages} {1877} (\bibinfo {year} {1998})}\BibitemShut {NoStop}%
\bibitem [{\citenamefont {Liang}\ \emph {et~al.}(2004)\citenamefont {Liang},
  \citenamefont {Lee}, \citenamefont {Resing}, \citenamefont {Ahn},\ and\
  \citenamefont {Klinman}}]{Liang2004}%
  \BibitemOpen
  \bibfield  {author} {\bibinfo {author} {\bibfnamefont {Z.-X.}\ \bibnamefont
  {Liang}}, \bibinfo {author} {\bibfnamefont {T.}~\bibnamefont {Lee}}, \bibinfo
  {author} {\bibfnamefont {K.~A.}\ \bibnamefont {Resing}}, \bibinfo {author}
  {\bibfnamefont {N.~G.}\ \bibnamefont {Ahn}}, \ and\ \bibinfo {author}
  {\bibfnamefont {J.~P.}\ \bibnamefont {Klinman}},\ }\href@noop {} {\bibfield
  {journal} {\bibinfo  {journal} {Proceedings of the National Academy of
  Sciences}\ }\textbf {\bibinfo {volume} {101}},\ \bibinfo {pages} {9556}
  (\bibinfo {year} {2004})}\BibitemShut {NoStop}%
\bibitem [{\citenamefont {Basran}\ \emph {et~al.}(1999)\citenamefont {Basran},
  \citenamefont {Sutcliffe},\ and\ \citenamefont {Scrutton}}]{Basran1999}%
  \BibitemOpen
  \bibfield  {author} {\bibinfo {author} {\bibfnamefont {J.}~\bibnamefont
  {Basran}}, \bibinfo {author} {\bibfnamefont {M.~J.}\ \bibnamefont
  {Sutcliffe}}, \ and\ \bibinfo {author} {\bibfnamefont {N.~S.}\ \bibnamefont
  {Scrutton}},\ }\href@noop {} {\bibfield  {journal} {\bibinfo  {journal}
  {Biochemistry}\ }\textbf {\bibinfo {volume} {38}},\ \bibinfo {pages} {3218}
  (\bibinfo {year} {1999})}\BibitemShut {NoStop}%
\bibitem [{\citenamefont {Knapp}\ \emph {et~al.}(2002)\citenamefont {Knapp},
  \citenamefont {Rickert},\ and\ \citenamefont {Klinman}}]{Knapp2002}%
  \BibitemOpen
  \bibfield  {author} {\bibinfo {author} {\bibfnamefont {M.~J.}\ \bibnamefont
  {Knapp}}, \bibinfo {author} {\bibfnamefont {K.}~\bibnamefont {Rickert}}, \
  and\ \bibinfo {author} {\bibfnamefont {J.~P.}\ \bibnamefont {Klinman}},\
  }\href@noop {} {\bibfield  {journal} {\bibinfo  {journal} {Journal of the
  American Chemical Society}\ }\textbf {\bibinfo {volume} {124}},\ \bibinfo
  {pages} {3865} (\bibinfo {year} {2002})}\BibitemShut {NoStop}%
\bibitem [{\citenamefont {Cha}\ \emph {et~al.}(1989)\citenamefont {Cha},
  \citenamefont {Murray},\ and\ \citenamefont {Klinman}}]{Cha1989}%
  \BibitemOpen
  \bibfield  {author} {\bibinfo {author} {\bibfnamefont {Y.}~\bibnamefont
  {Cha}}, \bibinfo {author} {\bibfnamefont {C.~J.}\ \bibnamefont {Murray}}, \
  and\ \bibinfo {author} {\bibfnamefont {J.~P.}\ \bibnamefont {Klinman}},\
  }\href@noop {} {\bibfield  {journal} {\bibinfo  {journal} {Science}\ }\textbf
  {\bibinfo {volume} {243}},\ \bibinfo {pages} {1325} (\bibinfo {year}
  {1989})}\BibitemShut {NoStop}%
\bibitem [{\citenamefont {Klinman}\ and\ \citenamefont
  {Kohen}(2013)}]{Klinman2013}%
  \BibitemOpen
  \bibfield  {author} {\bibinfo {author} {\bibfnamefont {J.~P.}\ \bibnamefont
  {Klinman}}\ and\ \bibinfo {author} {\bibfnamefont {A.}~\bibnamefont
  {Kohen}},\ }\href@noop {} {\bibfield  {journal} {\bibinfo  {journal} {Annual
  Review of Biochemistry}\ }\textbf {\bibinfo {volume} {82}},\ \bibinfo {pages}
  {471} (\bibinfo {year} {2013})}\BibitemShut {NoStop}%
\bibitem [{\citenamefont {Sacquin-Mora}\ \emph {et~al.}(2007)\citenamefont
  {Sacquin-Mora}, \citenamefont {Laforet},\ and\ \citenamefont
  {Lavery}}]{Sacquin-Mora2007}%
  \BibitemOpen
  \bibfield  {author} {\bibinfo {author} {\bibfnamefont {S.}~\bibnamefont
  {Sacquin-Mora}}, \bibinfo {author} {\bibfnamefont {E.}~\bibnamefont
  {Laforet}}, \ and\ \bibinfo {author} {\bibfnamefont {R.}~\bibnamefont
  {Lavery}},\ }\href@noop {} {\bibfield  {journal} {\bibinfo  {journal}
  {Proteins: Structure, Function, and Bioinformatics}\ }\textbf {\bibinfo
  {volume} {67}},\ \bibinfo {pages} {350} (\bibinfo {year} {2007})}\BibitemShut
  {NoStop}%
\bibitem [{\citenamefont {Juanico}\ \emph {et~al.}(2007)\citenamefont
  {Juanico}, \citenamefont {Sanejouand}, \citenamefont {Piazza},\ and\
  \citenamefont {De~Los~Rios}}]{Juanico2007}%
  \BibitemOpen
  \bibfield  {author} {\bibinfo {author} {\bibfnamefont {B.}~\bibnamefont
  {Juanico}}, \bibinfo {author} {\bibfnamefont {Y.-H.}\ \bibnamefont
  {Sanejouand}}, \bibinfo {author} {\bibfnamefont {F.}~\bibnamefont {Piazza}},
  \ and\ \bibinfo {author} {\bibfnamefont {P.}~\bibnamefont {De~Los~Rios}},\
  }\href@noop {} {\bibfield  {journal} {\bibinfo  {journal} {Phys. Rev. Lett.}\
  }\textbf {\bibinfo {volume} {99}},\ \bibinfo {pages} {238104} (\bibinfo
  {year} {2007})}\BibitemShut {NoStop}%
\bibitem [{\citenamefont {Aubailly}\ and\ \citenamefont
  {Piazza}(2015)}]{Aubailly2015}%
  \BibitemOpen
  \bibfield  {author} {\bibinfo {author} {\bibfnamefont {S.}~\bibnamefont
  {Aubailly}}\ and\ \bibinfo {author} {\bibfnamefont {F.}~\bibnamefont
  {Piazza}},\ }\href@noop {} {\bibfield  {journal} {\bibinfo  {journal}
  {Scientific Reports}\ }\textbf {\bibinfo {volume} {5}},\ \bibinfo {pages}
  {14874} (\bibinfo {year} {2015})}\BibitemShut {NoStop}%
\bibitem [{\citenamefont {Kamal}\ \emph {et~al.}(2012)\citenamefont {Kamal},
  \citenamefont {Mohammad}, \citenamefont {Krishnamoorthy},\ and\ \citenamefont
  {Rao}}]{Kamal:2012aa}%
  \BibitemOpen
  \bibfield  {author} {\bibinfo {author} {\bibfnamefont {M.~Z.}\ \bibnamefont
  {Kamal}}, \bibinfo {author} {\bibfnamefont {T.~A.~S.}\ \bibnamefont
  {Mohammad}}, \bibinfo {author} {\bibfnamefont {G.}~\bibnamefont
  {Krishnamoorthy}}, \ and\ \bibinfo {author} {\bibfnamefont {N.~M.}\
  \bibnamefont {Rao}},\ }\href@noop {} {\bibfield  {journal} {\bibinfo
  {journal} {Plos One}\ }\textbf {\bibinfo {volume} {7}},\ \bibinfo {pages} {1}
  (\bibinfo {year} {2012})}\BibitemShut {NoStop}%
\bibitem [{\citenamefont {Guo}\ \emph {et~al.}(2012)\citenamefont {Guo},
  \citenamefont {He}, \citenamefont {Huang}, \citenamefont {Liu}, \citenamefont
  {Liu},\ and\ \citenamefont {Yang}}]{Guo:2012aa}%
  \BibitemOpen
  \bibfield  {author} {\bibinfo {author} {\bibfnamefont {X.}~\bibnamefont
  {Guo}}, \bibinfo {author} {\bibfnamefont {D.}~\bibnamefont {He}}, \bibinfo
  {author} {\bibfnamefont {L.}~\bibnamefont {Huang}}, \bibinfo {author}
  {\bibfnamefont {L.}~\bibnamefont {Liu}}, \bibinfo {author} {\bibfnamefont
  {L.}~\bibnamefont {Liu}}, \ and\ \bibinfo {author} {\bibfnamefont
  {H.}~\bibnamefont {Yang}},\ }\href@noop {} {\bibfield  {journal} {\bibinfo
  {journal} {Computational and Theoretical Chemistry}\ }\textbf {\bibinfo
  {volume} {995}},\ \bibinfo {pages} {17 } (\bibinfo {year}
  {2012})}\BibitemShut {NoStop}%
\bibitem [{\citenamefont {Atilgan}\ \emph {et~al.}(2001)\citenamefont
  {Atilgan}, \citenamefont {Durell}, \citenamefont {Jernigan}, \citenamefont
  {Demirel}, \citenamefont {Keskin},\ and\ \citenamefont
  {Bahar}}]{Atilgan2001}%
  \BibitemOpen
  \bibfield  {author} {\bibinfo {author} {\bibfnamefont {A.}~\bibnamefont
  {Atilgan}}, \bibinfo {author} {\bibfnamefont {S.}~\bibnamefont {Durell}},
  \bibinfo {author} {\bibfnamefont {R.}~\bibnamefont {Jernigan}}, \bibinfo
  {author} {\bibfnamefont {M.}~\bibnamefont {Demirel}}, \bibinfo {author}
  {\bibfnamefont {O.}~\bibnamefont {Keskin}}, \ and\ \bibinfo {author}
  {\bibfnamefont {I.}~\bibnamefont {Bahar}},\ }\href {\doibase
  10.1016/S0006-3495(01)76033-X} {\bibfield  {journal} {\bibinfo  {journal}
  {Biophysical Journal}\ }\textbf {\bibinfo {volume} {80}},\ \bibinfo {pages}
  {505} (\bibinfo {year} {2001})}\BibitemShut {NoStop}%
\bibitem [{\citenamefont {Tirion}(1996)}]{Tirion1996}%
  \BibitemOpen
  \bibfield  {author} {\bibinfo {author} {\bibfnamefont {M.~M.}\ \bibnamefont
  {Tirion}},\ }\href {\doibase 10.1103/PhysRevLett.77.1905} {\bibfield
  {journal} {\bibinfo  {journal} {Phys. Rev. Lett.}\ }\textbf {\bibinfo
  {volume} {77}},\ \bibinfo {pages} {1905} (\bibinfo {year}
  {1996})}\BibitemShut {NoStop}%
\bibitem [{\citenamefont {Bahar}\ and\ \citenamefont {Cui}(2005)}]{Bahar2005}%
  \BibitemOpen
  \bibfield  {author} {\bibinfo {author} {\bibfnamefont {I.}~\bibnamefont
  {Bahar}}\ and\ \bibinfo {author} {\bibfnamefont {Q.}~\bibnamefont {Cui}},\
  }\href@noop {} {\emph {\bibinfo {title} {Normal Mode Analysis: Theory and
  Applications to Biological and Chemical Systems}}},\ edited by\ \bibinfo
  {editor} {\bibfnamefont {B.~R.}\ \bibnamefont {CRC~Press}},\ \bibinfo
  {series} {Mathematical \& Computational Biology Series}, Vol.~\bibinfo
  {volume} {9}\ (\bibinfo  {publisher} {CRC Press},\ \bibinfo {year}
  {2005})\BibitemShut {NoStop}%
\bibitem [{\citenamefont {Filoche}\ and\ \citenamefont
  {Mayboroda}(2012)}]{Filoche2012}%
  \BibitemOpen
  \bibfield  {author} {\bibinfo {author} {\bibfnamefont {M.}~\bibnamefont
  {Filoche}}\ and\ \bibinfo {author} {\bibfnamefont {S.}~\bibnamefont
  {Mayboroda}},\ }\href@noop {} {\bibfield  {journal} {\bibinfo  {journal}
  {Proceedings of the National Academy of Sciences}\ }\textbf {\bibinfo
  {volume} {109}},\ \bibinfo {pages} {14761} (\bibinfo {year}
  {2012})}\BibitemShut {NoStop}%
\bibitem [{\citenamefont {Lefebvre}\ \emph {et~al.}(2016)\citenamefont
  {Lefebvre}, \citenamefont {Gondel}, \citenamefont {Dubois}, \citenamefont
  {Atlan}, \citenamefont {Feppon}, \citenamefont {Labb\'e}, \citenamefont
  {Gillot}, \citenamefont {Garelli}, \citenamefont {Ernoult}, \citenamefont
  {Mayboroda}, \citenamefont {Filoche},\ and\ \citenamefont
  {Sebbah}}]{Lefebvre2016}%
  \BibitemOpen
  \bibfield  {author} {\bibinfo {author} {\bibfnamefont {G.}~\bibnamefont
  {Lefebvre}}, \bibinfo {author} {\bibfnamefont {A.}~\bibnamefont {Gondel}},
  \bibinfo {author} {\bibfnamefont {M.}~\bibnamefont {Dubois}}, \bibinfo
  {author} {\bibfnamefont {M.}~\bibnamefont {Atlan}}, \bibinfo {author}
  {\bibfnamefont {F.}~\bibnamefont {Feppon}}, \bibinfo {author} {\bibfnamefont
  {A.}~\bibnamefont {Labb\'e}}, \bibinfo {author} {\bibfnamefont
  {C.}~\bibnamefont {Gillot}}, \bibinfo {author} {\bibfnamefont
  {A.}~\bibnamefont {Garelli}}, \bibinfo {author} {\bibfnamefont
  {M.}~\bibnamefont {Ernoult}}, \bibinfo {author} {\bibfnamefont
  {S.}~\bibnamefont {Mayboroda}}, \bibinfo {author} {\bibfnamefont
  {M.}~\bibnamefont {Filoche}}, \ and\ \bibinfo {author} {\bibfnamefont
  {P.}~\bibnamefont {Sebbah}},\ }\href@noop {} {\bibfield  {journal} {\bibinfo
  {journal} {Phys. Rev. Lett.}\ }\textbf {\bibinfo {volume} {117}},\ \bibinfo
  {pages} {074301} (\bibinfo {year} {2016})}\BibitemShut {NoStop}%
\bibitem [{\citenamefont {Arnold}\ \emph {et~al.}(2018)\citenamefont {Arnold},
  \citenamefont {David}, \citenamefont {Filoche}, \citenamefont {Jerison},\
  and\ \citenamefont {Mayboroda}}]{Arnold2018}%
  \BibitemOpen
  \bibfield  {author} {\bibinfo {author} {\bibfnamefont {D.~N.}\ \bibnamefont
  {Arnold}}, \bibinfo {author} {\bibfnamefont {G.}~\bibnamefont {David}},
  \bibinfo {author} {\bibfnamefont {M.}~\bibnamefont {Filoche}}, \bibinfo
  {author} {\bibfnamefont {D.}~\bibnamefont {Jerison}}, \ and\ \bibinfo
  {author} {\bibfnamefont {S.}~\bibnamefont {Mayboroda}},\ }\href@noop {}
  {\bibfield  {journal} {\bibinfo  {journal} {arXiv:1711.04888}\ } (\bibinfo
  {year} {2018})}\BibitemShut {NoStop}%
\bibitem [{\citenamefont {Yang}\ and\ \citenamefont {Bahar}(2005)}]{Yang2005}%
  \BibitemOpen
  \bibfield  {author} {\bibinfo {author} {\bibfnamefont {L.-W.}\ \bibnamefont
  {Yang}}\ and\ \bibinfo {author} {\bibfnamefont {I.}~\bibnamefont {Bahar}},\
  }\href {\doibase 10.1016/j.str.2005.03.015} {\bibfield  {journal} {\bibinfo
  {journal} {Structure}\ }\textbf {\bibinfo {volume} {13}},\ \bibinfo {pages}
  {893} (\bibinfo {year} {2005})}\BibitemShut {NoStop}%
\bibitem [{\citenamefont {Lyra}\ \emph {et~al.}(2015)\citenamefont {Lyra},
  \citenamefont {Mayboroda},\ and\ \citenamefont {Filoche}}]{Lyra2015}%
  \BibitemOpen
  \bibfield  {author} {\bibinfo {author} {\bibfnamefont {M.~L.}\ \bibnamefont
  {Lyra}}, \bibinfo {author} {\bibfnamefont {S.}~\bibnamefont {Mayboroda}}, \
  and\ \bibinfo {author} {\bibfnamefont {M.}~\bibnamefont {Filoche}},\
  }\href@noop {} {\bibfield  {journal} {\bibinfo  {journal} {EPL (Europhysics
  Letters)}\ }\textbf {\bibinfo {volume} {109}},\ \bibinfo {pages} {47001}
  (\bibinfo {year} {2015})}\BibitemShut {NoStop}%
\bibitem [{\citenamefont {Yan}\ \emph {et~al.}(2018)\citenamefont {Yan},
  \citenamefont {Ravasio}, \citenamefont {Brito},\ and\ \citenamefont
  {Wyart}}]{Yan2019}%
  \BibitemOpen
  \bibfield  {author} {\bibinfo {author} {\bibfnamefont {L.}~\bibnamefont
  {Yan}}, \bibinfo {author} {\bibfnamefont {R.}~\bibnamefont {Ravasio}},
  \bibinfo {author} {\bibfnamefont {C.}~\bibnamefont {Brito}}, \ and\ \bibinfo
  {author} {\bibfnamefont {M.}~\bibnamefont {Wyart}},\ }\href {\doibase
  10.1016/j.bpj.2018.05.015} {\bibfield  {journal} {\bibinfo  {journal}
  {Biophysical Journal}\ }\textbf {\bibinfo {volume} {114}},\ \bibinfo {pages}
  {2787} (\bibinfo {year} {2018})}\BibitemShut {NoStop}%
\bibitem [{\citenamefont {Porter}\ \emph {et~al.}(2004)\citenamefont {Porter},
  \citenamefont {Bartlett},\ and\ \citenamefont {Thornton}}]{Porter2004}%
  \BibitemOpen
  \bibfield  {author} {\bibinfo {author} {\bibfnamefont {C.~T.}\ \bibnamefont
  {Porter}}, \bibinfo {author} {\bibfnamefont {G.~J.}\ \bibnamefont
  {Bartlett}}, \ and\ \bibinfo {author} {\bibfnamefont {J.~M.}\ \bibnamefont
  {Thornton}},\ }\href {\doibase 10.1093/nar/gkh028} {\bibfield  {journal}
  {\bibinfo  {journal} {Nucleic Acids Research}\ }\textbf {\bibinfo {volume}
  {32}},\ \bibinfo {pages} {D129} (\bibinfo {year} {2004})}\BibitemShut
  {NoStop}%
\bibitem [{\citenamefont {Meyer}\ and\ \citenamefont
  {Klinman}(2005)}]{Meyer2005}%
  \BibitemOpen
  \bibfield  {author} {\bibinfo {author} {\bibfnamefont {M.~P.}\ \bibnamefont
  {Meyer}}\ and\ \bibinfo {author} {\bibfnamefont {J.~P.}\ \bibnamefont
  {Klinman}},\ }\href@noop {} {\bibfield  {journal} {\bibinfo  {journal}
  {Chemical Physics}\ }\textbf {\bibinfo {volume} {319}},\ \bibinfo {pages}
  {283} (\bibinfo {year} {2005})}\BibitemShut {NoStop}%
\bibitem [{\citenamefont {Piazza}(2014)}]{Piazza:2014aa}%
  \BibitemOpen
  \bibfield  {author} {\bibinfo {author} {\bibfnamefont {F.}~\bibnamefont
  {Piazza}},\ }\href@noop {} {\bibfield  {journal} {\bibinfo  {journal}
  {Physical Biology}\ }\textbf {\bibinfo {volume} {11}},\ \bibinfo {pages}
  {036003} (\bibinfo {year} {2014})}\BibitemShut {NoStop}%
\bibitem [{\citenamefont {Schwartz}\ and\ \citenamefont
  {Schramm}(2009)}]{Schwartz2009}%
  \BibitemOpen
  \bibfield  {author} {\bibinfo {author} {\bibfnamefont {S.~D.}\ \bibnamefont
  {Schwartz}}\ and\ \bibinfo {author} {\bibfnamefont {V.~L.}\ \bibnamefont
  {Schramm}},\ }\href@noop {} {\bibfield  {journal} {\bibinfo  {journal}
  {Nature Chemical Biology}\ }\textbf {\bibinfo {volume} {5}},\ \bibinfo
  {pages} {551} (\bibinfo {year} {2009})}\BibitemShut {NoStop}%
\end{thebibliography}

%

\end{document}